\documentclass{ornltm}

\usepackage{threeparttable} 
\usepackage{booktabs} 
\usepackage[final]{microtype} 
\usepackage{multirow} 

\usepackage{subcaption}
\usepackage{caption}
\usepackage{gensymb}
\usepackage{todonotes}
\usepackage{comment}

\usepackage{biblatex}  
\bibliography{mpex.bib} 
\definecolor{note}{RGB}{48,96,192}

\author{ORNL \and Gary Staebler \and  Rhea Barnett \and Mark Cianciosa \and Rinkle Juneja \and Atul Kumar \and Wouter Tierens \and Minglei Yang \and Cory Hauck \and Richard Archibald \and Viktor Reshniak \and Pablo Seleson \and Sam Reeve \and Gregory Watson \and John Duggan \and LLNL \and Ben Dudson \and Vasily Geyko}

\title{MPEX AI Digital Twins Milestone Report}

\date{\today}

\reportnum{}

\division{Fusion Energy Division}

\begin{document}
\frontmatter
\tableofcontents

\mainmatter
\section{EXECUTIVE SUMMARY}
This is the six month progress report to Fusion Energy Science (FES) and the American Science Cloud (AmSC) on the MPEX AI Digtial Twins project that was started in October 2025. There are two milestones to demonstrate the Artificial Intelligence (AI) advantage for MPEX operations and scientific discovery, that will be completed by June 2026. The first is a Helicon AI Hot-Spot Controller (Sec.~\ref{sec:AI-Controller}), which is the helicon heating component of the more comprehensive planned MPEX AI Hot Spot Digital Twin (Sec.~\ref{sec:Hot-Spot}). The second is an E-beam Damage Assessment Digital Twin (Sec.~\ref{sec:mat:ai}), which is a reduced electron beam damage modality prototype for the MPEX AI Damage Assessment Digital Twin (Sec.~\ref{sec:mat}). These two phase I milestones are on track for the June demonstration. 

In addition to these two milestones, progress on configuring the Galaxy software interface for automation, validation and data analysis is reported (Sec.~\ref{sec:workflows}). This interface now connects a subset of the main physics simulation codes to DOE HPC resources and will connect to the MPEX data acquisition system so that analysis of data, validation and execution of simulations can be performed by the scientist or by AI-Agents. When AmSC is ready to accept connections and data, Galaxy will be the MPEX interface to AmSC. 

\subsection{Progress on the Helicon AI Hot Spot Controller}
A physics fidelity hierarchy is used in generating the training data for the AI surrogates and Digital Twins. Our highest fidelity model of the helicon wave physics is the COMSOL code. For the first time, 3D COMSOL simulations of the MPEX helicon system were made. Even though the background plasma was assumed to be uniform on magnetic flux surfaces, the helicon absorption pattern was computed to be localized in the 2D plane (radial and azimuthal) perpendicular to the (axial) magnetic field lines. Using the 3D COMSOL heating source in the high-fidelity plasma turbulence simulation code Hermes-3 showed that the plasma generation follows the helicon heating pattern with only weak turbulent transport across magnetic fields. These results nicely reproduce the localized nature of the hot spots observed on the MPEX target. This result motivated revising the reduced plasma model used to train the AI Hot-Spot controller from a flux surface averaged model to a 3D flux tube model. Variations in the magnetic field line projection from the target to the helicon window, induced by changes in coil currents, were used to train the Generative AI Hot Spot model. It was found that varying the coil currents was effective at spreading out the heat to the window in the direction of the magnetic field (axial) but that in the azimuthal direction along the window, the heat flux map remains asymmetric due to the azimuthal variation in the plasma parallel heat flux. Because of this the AI-agentic optimization was able to minimize the axial heat asymmetry but not the azimuthal heat asymmetry by changing coil current. Experimental data from proto-MPEX shows that, at sufficiently high helicon power and plasma density, the helicon heating is centrally peaked. The azimuthal variation of the plasma is much lower in this central heating regime.

For the June Demonstration, new experimental data from proto-MPEX-lite operations in progress will be used to train the Helicon AI Hot Spot Controller to find the optimum power, neutral gas pressure, and coil currents required to obtain central plasma heating. Central plasma heating has been found to be the most critical factor in reducing azimuthal hot spots. The reduced plasma model will be made more realistic by including flux tube parallel collisional transport and fluid neutral recycling using the exiting C1 code. This will allow for temperature differences between the helicon window region and the target, or back end dump plate, making the reconstruction of the helicon heating pattern in the plasma more accurate.

\subsection{Progress on the E-beam Damage Assessment Digital Twin}
Two new AI models have been developed for E-beam damage assessment of tungsten-based plasma facing component candidate materials: automated crack identification from experimental characterization, and crack damage prediction combining this automated crack image analysis, experimental conditions, and material properties. The first tool analyzes microscopy images of experimentally damaged e-beam samples, automating information extraction across characterization modes and scales, and through an image processing pipeline identifies all cracks in each image. Reliable automation results are obtained across both highly cracked and undamaged surfaces and an initial version of this tool has been deployed through Galaxy. The second tool predicts crack density given experimental base temperature and heat flux for a given tungsten alloy of interest. This AI vision transformer model was trained through patch-based analysis of the processed crack network images and embedded in a feature space in order to generalize and enable future extensibility. Experimental process conditions across tungsten alloy chemistries and microstructures were then used to learn similarity distance functions in this embedding space in order to combine all available information in a consistent model. Predictions of crack density using a kernel regressor show reasonably good prediction accuracy where data is available; however, the available data is extremely sparse.  Physics fracture modeling has been developed to support filling this process parameter space. The CabanaPD mechanics simulation tool has been extended to anisotropic continuum mechanics necessary for the tungsten alloys of interest and has been successfully validated for a range of mechanical responses. Direct computational experiments of the same E-beam conditions have been demonstrated and are now ready to support AI model accuracy improvement by extending the training set.

The June Demonstration will couple these efforts into a simulation-enabled AI digital twin for E-beam crack damage prediction. Extension from the current work first includes focusing the AI material assessment on learning the boundaries between regimes of damage (crack networks, isolated damage, surface roughness, no modification, ...). In addition, a sampling approach is currently being implemented to automate crack prediction simulations to fill gaps in the available experimental results. All AI and physics tools will be coupled through the Galaxy interface, run on DOE HPC, and made available through the web interface. This demonstration will thus predict the damage regimes of the tungsten alloys of interest to exemplify future MPEX use for candidate material assessment. In addition, the AI-based automated crack identification will be independently demonstrated to highlight intended use within the MPEX Surface Analysis Station (SAS).

\subsection{Phase II Genesis Mission RFA}
Future plans for the MPEX AI Digital Twins project will include new digital twins for control of the MPEX whole facility heating and cooling systems. This will enable operation with AI-agentic decisions to avoid shutdown during the extremely long pulse length of MPEX. The ICH and ECH heating systems will be added to the MPEX AI Hot-Spot Controller. Physics modeling of other modalities of plasma material damage (erosion, redeposition, sputtering,..) will be expanded to train the MPEX AI Material Damage Assessment Digital Twin. The GALAXY framework will add communication with the MPEX data acquisition system and the AmSC so that MPEX will be a fully functioning science platform when it begins operations. Details can be found in the phase II Genesis Mission RFA application to be submitted by May 19th. 

\section{INTRODUCTION AND BACKGROUND}


\begin{figure}
  \centering
  \includegraphics[scale=1.0]{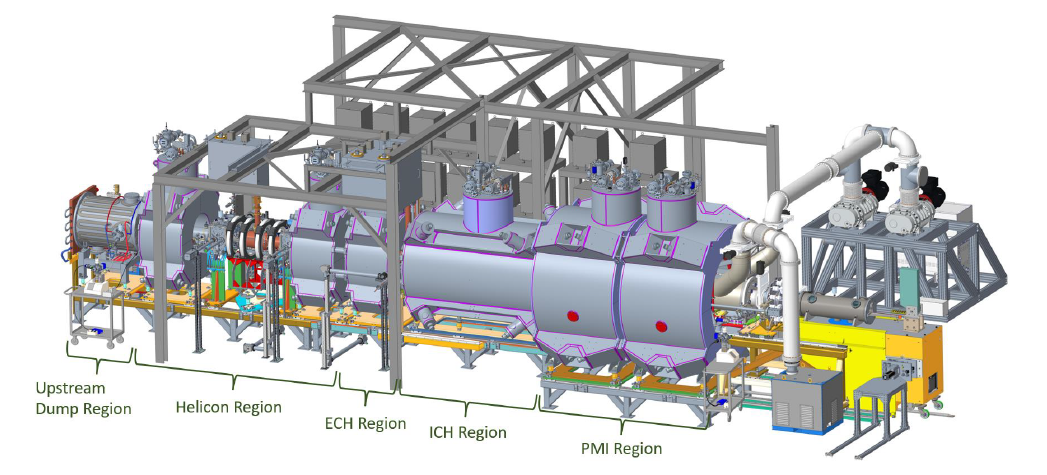}
  \caption{Drawing of the Material Plasma Exposure eXperiment (MPEX) showing the Helicon plasma source, Electron Cyclotron Heating (ECH), Ion Cyclotron Heating (ICH) and Plasma Material Interaction (PMI) regions }
\label{fig:MPEX}
\end{figure}

The MPEX device will begin commissioning at the end of FY26. A smaller proto-MPEX was operated for 14,666 plasma discharges, providing the data for our initial AI Hot Spot Controller training progress. A smaller machine with only helicon heating, proto-MPEX-lite, began operation in September of 2025, to test a new window for the helicon plasma source. Proto-MPEX-lite is now providing useful data for this project. An additional IR camera view of the rear dump plate was installed that will improve the Helicon AI Hot Spot Controller training.  

The scientific mission of MPEX is to qualify materials of different composition for use in the high energy and plasma flux conditions of a fusion power plant. The materials exposed in MPEX will in some cases be exposed to high neutron fluxes at other ORNL facilities to measure the changes to their PMI properties. The targets exposed in MPEX will be transported under vacuum to a Surface Analysis Station (SAS). The SAS will be equipped with the following diagnostics: focused ion beam (FIB) for trench milling, 100-400 angstrom resolution scanning electron microscope (SEM), surface mapping x-ray spectrometer, high resolution camera, and a future upgrade to a laser induced breakdown spectroscopy quadruple mass spectrometer (LIBS-QMS). The MPEX experiments will generate diverse pre- and post-exposure measurement data of detailed material properties down to the microstructure in 3D for post-exposure assessment of PMI damage (e.g. cracking, sputtering, erosion and redeposition of the material). Physics models for the PMI, and how the material composition and manufacturing impact its performance under high energy plasma exposure, need to be validated with MPEX data to guide the selection of new candidate materials. The MPEX AI Damage Assessment Digital Twin developed in this project will enable automated characterization and prediction of material behavior for MPEX target experiments.

The \textit{MPEX AI Digital Twins} project AI advantage is to train AI models from experimental and physics simulation data in order to process data and provide analysis for PMI, control and improve the operations of the MPEX device, and maximize the scientific output of MPEX. Ultimately, an AI digital twin of MPEX material assessment metrics for tested and synthetic material types with simulated PMI will be trained by the AI Modeling Teams on the experimental and physics simulation data submitted to the American Science Cloud by this project. A purely empirical search for the best material is inefficient given the finite number of samples that can be tested on MPEX. In order to expand the material properties database for training the MPEX Damage Assessment AI Digital Twin, and to gain physics understanding of the PMI processes, physics models of the material properties and PMI processes are required. The physics simulations provide detailed simulation data, like impact angles for plasma ions, sputtering yields, transport of the ionized sputtered target material in the plasma, and redeposition locations. This simulation data expands the measurement data for deeper physics understanding. The experimental data is essential to validate the PMI and material structure simulation models. The validated models can then be used to generate new simulation data of MPEX material assessments for synthetic material compositions that have not been exposed in MPEX. These predictive simulations, plus the whole experimental dataset, will be used to train the MPEX AI Damage Assessment Digital Twin allowing a rapid generative AI search for new materials with reduced PMI damage by interpolating the domain of the training set. These new optimum materials can be simulated with the physics codes and/or tested in MPEX. The ability of generative AI methods to interpolate multi-dimensional parameter spaces and generate virtual data is exploited for a more efficient search for optimum materials. 

We next describe the six-month progress for both primary thrusts of the project, the MPEX AI Hot Spot Controller Digital Twin and MPEX AI Damage Assessment Digital Twin, as well as the automation and workflow efforts. Each thrust details both AI development and physics modeling to support the AI efforts. Next, we describe continuing work towards the June milestone demonstrations for each digital twin. Finally, we conclude with future work across the project.

\clearpage
\section{MPEX AI HOT SPOT DIGITAL TWIN}
\label{sec:Hot-Spot}
This six-month report summarizes progress on the development of a predictive plasma--material interaction (PMI) modeling framework for MPEX and its integration into an AI-enabled hot-spot control strategy. The central objective is to connect experimentally observed heat-flux localization (i.e. hot spots) at the \textit{helicon window} and the \textit{target} to the underlying magnetic topology, radio-frequency (RF) heating physics, and plasma and impurity transport, and to use this understanding to guide experimental control in real time.

The work is organized around three tightly coupled elements. First, Proto-MPEX infrared (IR) measurements provide direct evidence that surface heating at the helicon window is strongly localized and highly sensitive to magnetic topology, plasma density and heating power. While simultaneous window and target heat-flux measurements are not available for the same experimental conditions in proto-MPEX data, the existing data clearly demonstrate that localized heating is influenced by magnetic connectivity. Second, three-dimensional RF modeling shows that helicon power deposition is intrinsically azimuthally asymmetric and edge localized, such that only a subset of magnetic field lines carries enhanced energy. Third, a flux-tube heat-flux mapping framework is used to transform measured window heat flux into a field-line-resolved quantity that can be propagated to other surfaces and evaluated under modified magnetic configurations. Together, these elements establish the basis for predictive PMI modeling and AI-driven hot-spot control.

\subsection{Helicon AI Hot-Spot Controller}
\label{sec:AI-Controller}

In MPEX, plasma heat flux to the walls must be controlled to avoid localized hot spots that can crack antenna windows or damage other plasma-facing components. Among the controlling factors, the magnetic configuration, determined by the coil currents, plays a dominant role in shaping plasma trajectories, RF power absorption, and the resulting heat-flux distribution on material surfaces. Because MPEX operates with multiple heating systems in a strongly nonuniform magnetic geometry, localized heating can develop even when the total absorbed power remains within acceptable limits. The operational challenge is therefore not only to deliver power efficiently to the plasma, but also to tailor the magnetic topology so that excessive heat loads do not develop on vulnerable components. Hot Spots on the helicon window are of particular concern since they cause sputtering or could even crack the window. A new helicon window with coatings to reduce sputtering is now being tested in proto-MPEX-lite. In order to contribute to resolving the helicon window hot spot issue, a Helicon AI Hot-Spot Controller was chosen as the first milestone of this project.  

Manual optimization of coil currents is impractical due to the high dimensionality of the control space and the inherently nonlocal response of the system. A change in coil currents modifies the magnetic field, which in turn simultaneously affects RF heating, plasma transport, and the surface intercept of flux tubes. These coupled effects make intuitive or trial-and-error tuning inefficient.

To address this challenge, an AI-based controller is proposed to guide magnetic tuning, along with other operational parameters such as gas puffing and RF power, and to mitigate hot spots in real time. The approach combines experimental imaging, reduced-order physics models, and high-fidelity simulations within a control-oriented framework. In this formulation, the AI system does not replace the underlying physics; instead, it is trained using both experimental observations and physics-based synthetic data. This enables rapid prediction of heat-flux redistribution under changes in experimental control parameters and provides actionable guidance for controlling hot-spot formation during operation.

\subsection{Hot-Spot Identification and Control Objective}
\label{subsec:objective}

High-speed cameras provide real-time imaging of plasma-facing surfaces in MPEX and Proto-MPEX, capturing the onset, motion, and evolution of localized hot spots. A machine-learning model was trained to identify and analyze these events using convolutional neural networks for spatial feature extraction and temporal models for prediction and early warning.

The hot-spot analysis task includes both \textit{detection}, identifying the presence of potentially harmful localized heating, and \textit{characterization}, quantifying the location, spatial extent, intensity, and temporal evolution of the hot spot.

The initial database of hot-spot images from proto-MPEX was constructed by systematically choosing discharges with varying experimental control parameters, including coil currents, gas puffing, and RF power.

The control objective is to provide shot-to-shot or real-time recommendations for operating parameters that minimize hot spots on the helicon window and other critical surfaces while maintaining desired plasma flux to the target. The AI system is therefore designed not only to recognize hazardous heating patterns, but also to predict how these patterns evolve under changes in magnetic topology and other control inputs.

This capability is implemented through a physics-informed control loop:
\begin{enumerate}
    \item detect and localize hot spots from camera and diagnostic data,
    \item predict heat-flux redistribution under variations in control parameters,
    \item recommend operating conditions that reduce wall loading while preserving target heat flux and RF constraints.
\end{enumerate}

The effectiveness of this control strategy depends directly on the fidelity of the underlying simulation framework.

\subsubsection{Experimental Evidence of Localized Helicon Heating and Magnetic Topology Effects}
\label{subsec:expt}

\begin{figure}
    \centering
    \includegraphics[width=0.8\linewidth]{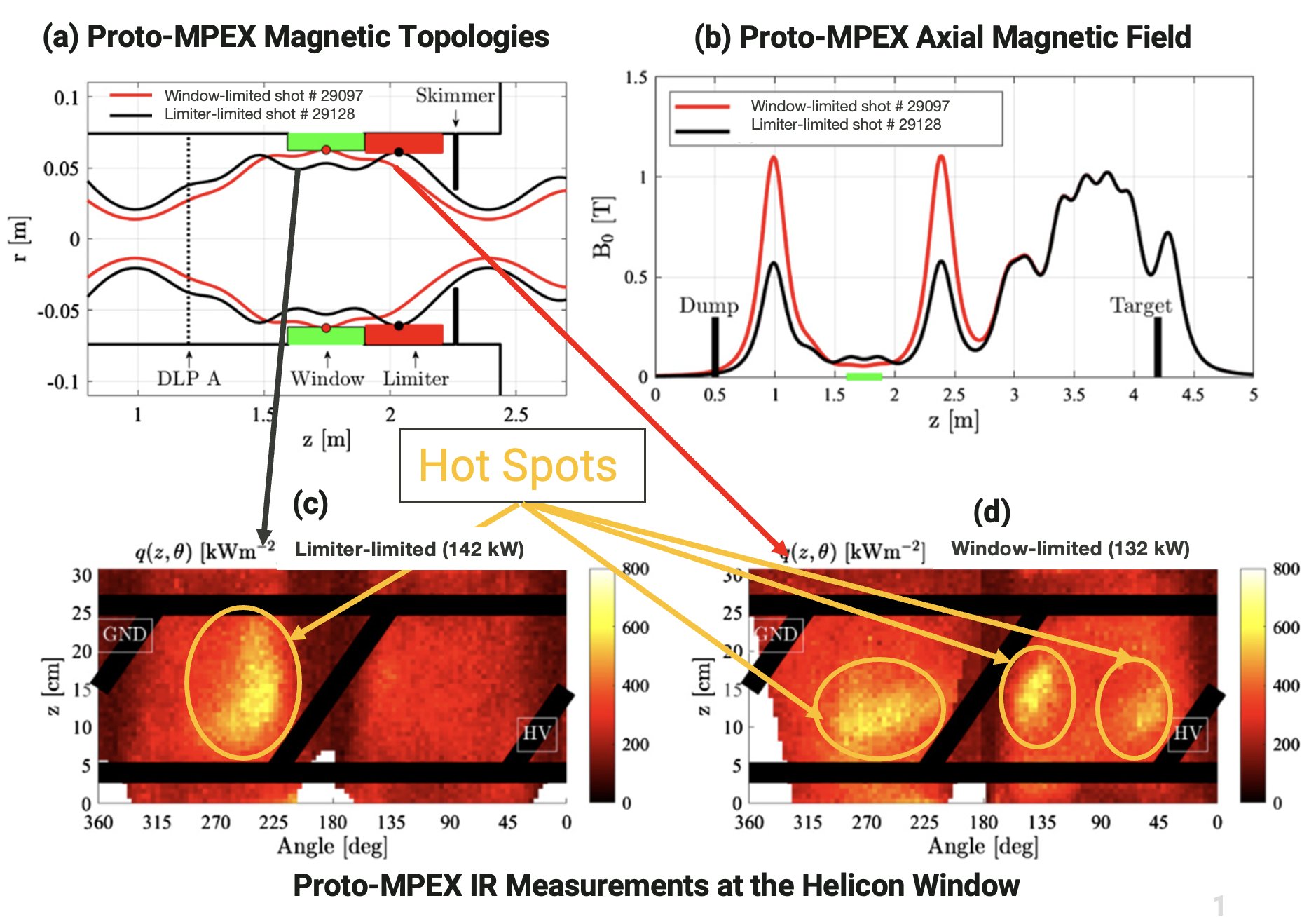}
    \caption{Proto-MPEX experimental observations of helicon-window heat flux. (a) Magnetic topology comparison between window-limited and limiter-limited configurations. (b) Corresponding axial magnetic field profiles. (c--d) IR heat-flux measurements showing localized hot spots in both axial and azimuthal directions.}
    \label{fig:proto_ir}
\end{figure}

\begin{figure}
    \centering
    \includegraphics[width=0.5\linewidth]{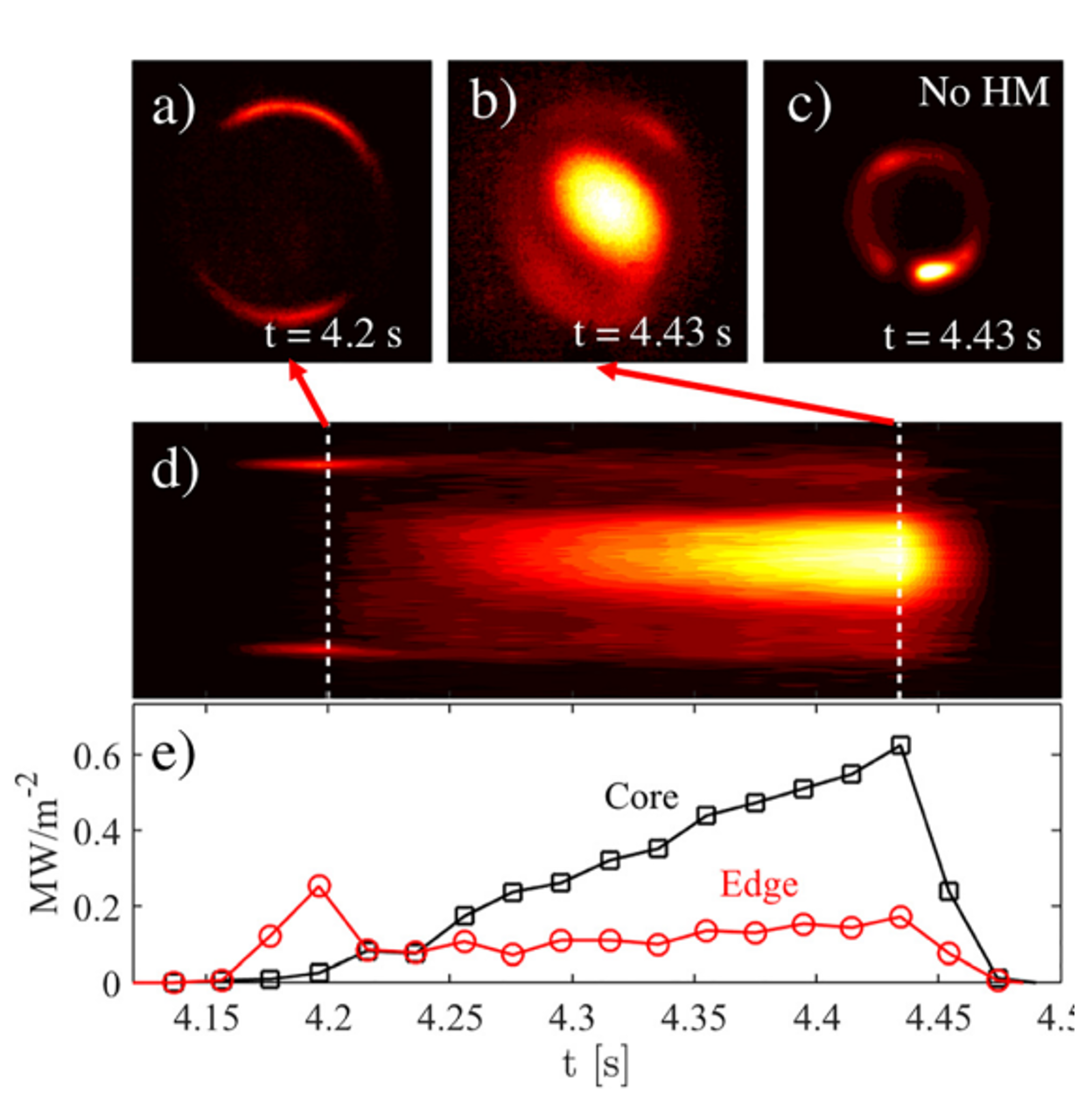}
    \caption{Proto-MPEX target heat-flux evolution from the same experimental shot as Fig.~\ref{fig:proto_ir}. The measurements show a transition from edge-localized ring heating to centrally peaked heating depending on RF power, gas puffing, and magnetic configuration.}
    \label{fig:proto_target}
\end{figure}

Infrared (IR) measurements from Proto-MPEX provide direct evidence that heat flux at both the helicon window and downstream target is strongly structured, with clear dependence on magnetic topology, RF power, and fueling conditions. figures~\ref{fig:proto_ir} and \ref{fig:proto_target} present helicon-window and target measurements obtained from \textit{similar experiments}, enabling direct comparison of upstream and downstream heat-flux behavior under identical operating conditions.

Two distinct magnetic topologies are examined. In the \textit{window-limited configuration}, magnetic field lines intersect the helicon window directly, resulting in strong localized heat deposition and pronounced azimuthal asymmetry. In the \textit{limiter-limited configuration}, field lines are intercepted downstream, reducing and redistributing the heat flux incident on the window.

Comparison of these configurations shows that transitioning from a window-limited to a limiter-limited topology reduces the heat flux to the helicon window by approximately \textit{30\%}. This demonstrates that surface heat loads are strongly governed by magnetic connectivity and not solely by total absorbed RF power.

The target measurements show that the heat-flux distribution can be significantly modified by varying RF power, gas puffing, and magnetic field configuration. Under certain conditions, heating is concentrated in a \textit{ring-like structure at the plasma edge}, corresponding to edge-dominated RF absorption. With increasing RF power and gas puffing, the plasma transitions toward a \textit{centrally peaked heating profile}, consistent with helicon-mode-dominated core heating.

These observations indicate that plasma heating in MPEX is not only asymmetric azimuthally, but also \textit{regime-dependent}, with the dominant heat-flux channel shifting between edge and core regions. This behavior reflects the interplay between RF wave absorption, plasma density profiles, neutral fueling, and magnetic topology.

Together, the window and target measurements establish that localized heat loads in MPEX are determined by \textit{field-line-resolved transport}, where specific flux-tube bundles carry enhanced energy from the heating region to material surfaces. This provides the experimental basis for the flux-tube heat-flux mapping framework developed in this work and detailed later in section \ref{subsec:fluxtubemapping}.

\subsubsection{Asymmetric Helicon Heating and Transition to 3D RF Modeling}
\label{subsec:comsol}

The azimuthally asymmetric and localized heating observed in Proto-MPEX IR measurements motivates a fundamental shift in modeling strategy within this project. Earlier analyses of helicon heating rely primarily on two-dimensional (2D) azimuthally symmetric RF models, which assume a single azimuthal mode ($m=1$) and predict smooth, bulk heating profiles. However, these models fail to reproduce the experimentally observed hot spots on the helicon window.

In this work, the modeling approach transitions to fully three-dimensional (3D) RF simulations using COMSOL in order to resolve the geometric and azimuthal complexity of the helicon antenna. This approach allows the azimuthal spectrum of RF fields to be determined self-consistently by the antenna geometry and plasma conditions, and therefore captures the inherently azimuthally asymmetric  nature of helicon heating in MPEX.

\begin{figure}
    \centering
    \includegraphics[width=0.8\linewidth]{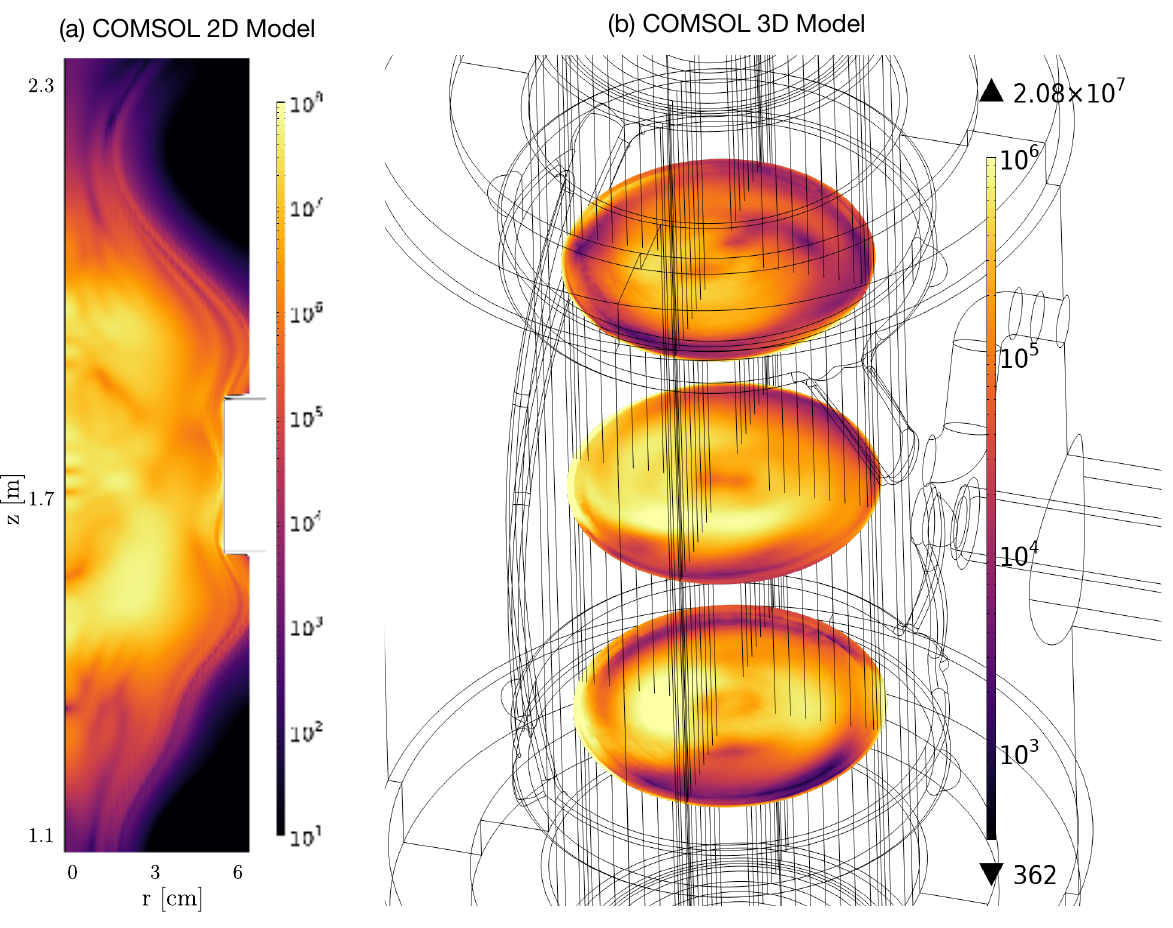}
    \caption{COMSOL RF heating simulations for Proto-MPEX. The 2D model shows diffuse, azimuthally symmetric heating dominated by bulk absorption, while the 3D model reveals discrete, azimuthally asymmetric, edge-localized RF power deposition aligned with magnetic field structures.}
    \label{fig:comsol_heating}
\end{figure}

The transition from 2D to 3D modeling reveals several key physical mechanisms governing helicon heating. The 3D simulations show finite \textit{edge-localized RF power deposition} near the plasma--window interface, in contrast to the bulk heating predicted by azimuthally symmetric models. This edge localization is consistent with IR observations of hot spots and provides a direct explanation for why heating appears at specific locations on the window rather than being broadly distributed across the plasma cross section.

The simulations also demonstrate the role of \textit{evanescent slow-wave excitation}. Although the slow wave remains evanescent under typical MPEX conditions, it is excited at the plasma boundary and generates localized electric fields near the window. These fields enhance RF absorption at the edge and contribute directly to localized heat loading.

In addition, the helicon antenna excites a \textit{broad azimuthal mode spectrum} rather than a single dominant mode. Higher-$m$ components are more radially localized and preferentially deposit energy near the plasma edge. This multi-mode structure cannot be captured within a 2D azimuthally symmetric approximation and is therefore essential for reproducing the observed heating patterns.

An important contribution arises from \textit{edge collisionality and neutral interactions}. The edge region in MPEX has elevated collisionality and significant neutral density due to recycling and gas fueling. Electron--neutral collisions, ionization, and charge-exchange processes enhance RF damping in this region, causing preferential deposition of RF power near the plasma boundary. As a result, the edge acts as a dominant sink for RF energy, further amplifying localized heating structures. The collisional heat conduction is strongly anistotropic, being much faster along magnetic field lines than perpendicular to them. This inhibits the azimuthal and radial spreading of the plasma away from the localized helicon heating regions. 

The simulations also reveal a \textit{core-to-edge heating transition} controlled by plasma profiles. Variations in density and temperature shift the dominant absorption region between the plasma core and edge. In particular, low edge density and edge-peaked temperature profiles favor strongly localized edge heating, while opposite to this leads to a more centrally peaked power deposition.

Taken together, these results show that helicon RF heating in MPEX is intrinsically three-dimensional and strongly influenced by edge physics. The combination of multi-mode RF excitation, edge collisionality, and neutral interactions produces discrete, flux-tube-aligned heating channels rather than uniform volumetric heating.

Within this project, this understanding further motivates the development of a flux-tube-based heat-flux mapping framework. The 3D COMSOL results indicate that only a subset of flux tubes receives enhanced RF power, which is then transported along magnetic field lines to plasma-facing components. This provides a direct physical link between RF heating, magnetic topology, and the localized hot spots observed experimentally.

\subsection{Plasma Modeling Components}

The MPEX modeling framework employs a multi-fidelity approach in which high-resolution plasma simulations, RF heating calculations, kinetic modeling, and reduced-order flux-tube models are coupled together. Each component captures a different aspect of the plasma--RF interaction physics, and their integration enables both predictive capability and computational efficiency.

\subsubsection{HERMES-3: High-Fidelity 3D Fluid Plasma Model}

HERMES-3 serves as the primary high-fidelity plasma model in the current framework. It is a three-dimensional fluid turbulence code that evolves plasma density, temperature, and flow fields in realistic magnetic geometries. Unlike simplified 1D or azimuthally symmetric models, HERMES-3 resolves turbulence-driven transport and captures 3D plasma structures that arise due to helicon heating and magnetic field shaping.

A key advantage of HERMES-3 is its wide-grid capability, which allows it to simulate large spatial domains while retaining resolution of critical edge physics. This is particularly important for MPEX, where plasma conditions vary strongly near material surfaces.

\begin{figure}
    \centering
    \includegraphics[width=\linewidth]{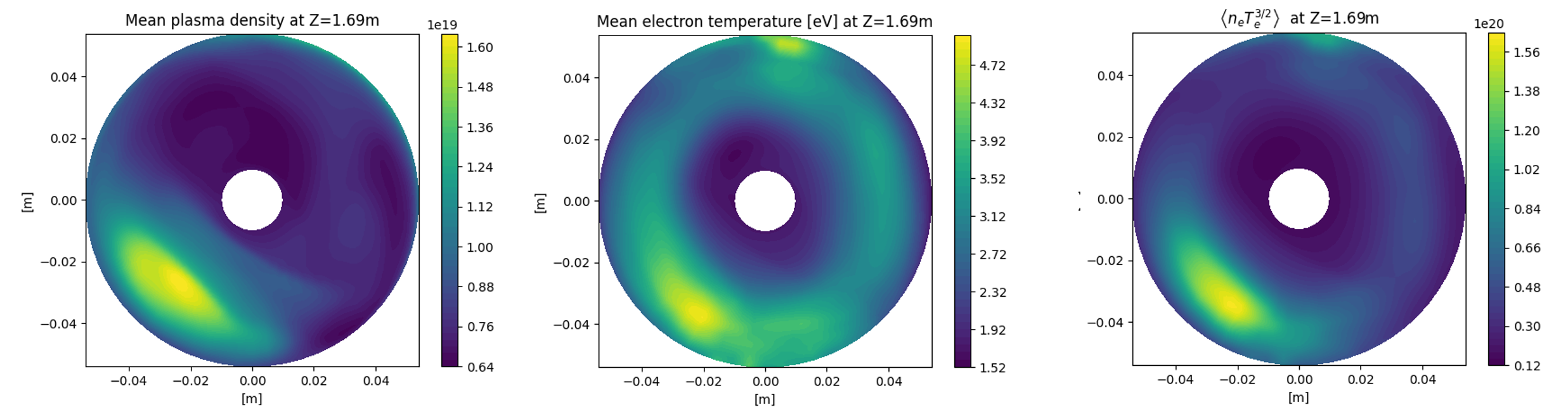}
    \caption{HERMES-3 time-averaged plasma profiles at a representative axial location. Left: plasma density $n_e$. Middle: electron temperature $T_e$. Right: proxy parallel heat flux quantity $\langle n_e T_e^{3/2} \rangle$. The maxima remain spatially localized and co-aligned across all quantities.}
    \label{fig:hermes_profiles}
\end{figure}

Fig.~\ref{fig:hermes_profiles} shows representative time-averaged plasma profiles obtained from HERMES-3. The density, temperature, and derived quantity $\langle n_e T_e^{3/2} \rangle$ exhibit strong spatial localization, reflecting the non-uniform heating and transport in the device. Importantly, the maxima of these quantities remain \textit{stationary in time} and are \textit{co-localized}, indicating that the dominant heating structures are persistent and not transient fluctuations. Turbulent eddies in the simulation drift around the plasma hot spot and only weakly spread out the plasma perpendicular to the magnetic field. 

These localized plasma features are consistent with the edge-localized and azimuthally asymmetric RF power deposition obtained from the 3D COMSOL simulations. In particular, the regions of enhanced $\langle n_e T_e^{3/2} \rangle$ align with the flux-tube bundles that receive strong RF heating, providing a direct connection between RF wave absorption, plasma response, and field-line-resolved heat transport.

\subsection{Flux-Tube Heat-Flux Mapping Framework}
\label{subsec:fluxtubemapping}

To connect RF heating and plasma transport to localized surface heat loads, a flux-tube-resolved heat-flux mapping framework has been developed. In this approach, the heat flux is represented as a parallel quantity,
\begin{equation}
Q_{\parallel}(l,m),
\end{equation}
defined on flux-tube identity coordinates corresponding to radial and azimuthal indices.

The key idea is to distinguish between the \emph{measurement surface} and the \emph{flux-tube identity space}. The helicon window is a cylindrical surface of fixed radius, so every physical heat-flux measurement at the window is recorded on the same cylindrical boundary. However, the field lines that intersect that cylinder originate from different radial and azimuthal locations upstream. To preserve this information, each flux tube is labeled by its source-space identity $(l,m)$, rather than by the fixed radius of the window itself.

The experimentally measured IR heat flux on the cylindrical window is used to reconstruct the upstream tube-carried quantity by accounting for magnetic-field incidence. At the window surface,
\begin{equation}
q_{\text{window}}(l,m) = Q_{\parallel}(l,m)\sin[\alpha_W(l,m)],
\end{equation}
where $\alpha_W$ is the local angle between the magnetic field and the window normal. This relation is inverted to determine $Q_{\parallel}(l,m)$ from the measured heat-flux map and the mapped flux-tube geometry.

Each flux tube is then traced through the magnetic field to determine its intersection with other material surfaces. The local surface heat flux is reconstructed using
\begin{equation}
q_{\text{surface}} = Q_{\parallel}\frac{|\mathbf{B}\cdot\hat{n}|}{|\mathbf{B}|},
\end{equation}
ensuring that heat flux remains attached to flux-tube identity while correctly accounting for surface geometry.

\begin{figure}[htbp]
    \centering
    \includegraphics[width=\linewidth]{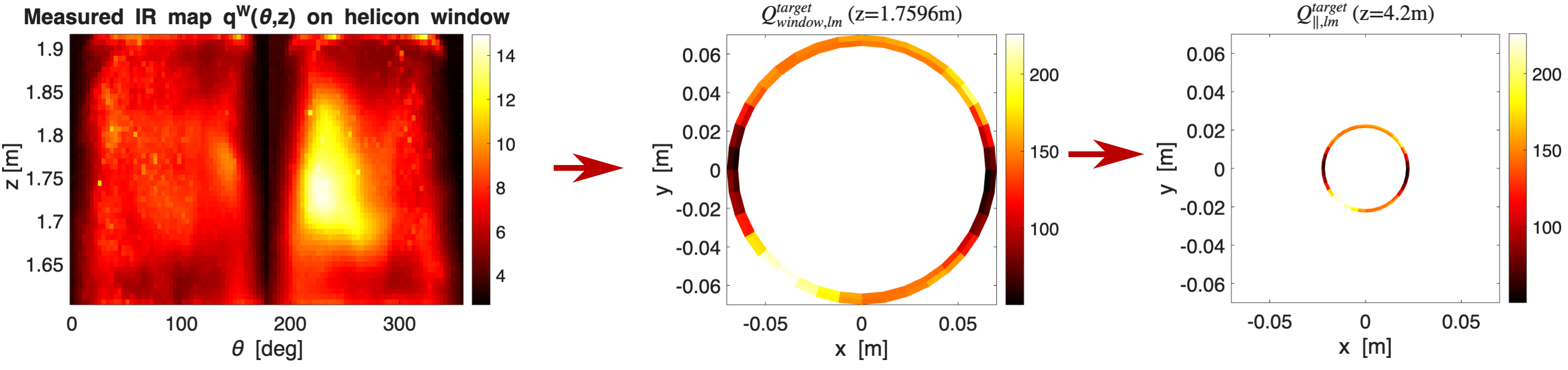}
    \caption{Flux-tube heat-flux mapping from Proto-MPEX measurements. Left: measured IR heat-flux map $q^W(\theta,z)$ on the helicon window. Middle: reconstructed flux-tube-resolved heat flux mapped to an annular $(r,\theta)$ representation at the window location. Right: predicted heat-flux distribution at the downstream target obtained by tracing the same flux tubes through the magnetic field.}
    \label{fig:flux_tube_mapping}
\end{figure}

Fig.~\ref{fig:flux_tube_mapping} illustrates the complete mapping process. The measured helicon-window heat flux is first converted into a flux-tube-resolved quantity, which is then propagated along magnetic field lines to predict heat-flux deposition at other surfaces. The figure demonstrates that localized hot spots observed at the window correspond to specific flux-tube bundles, which map to distinct regions at the target.

A major advantage of this framework is that it provides a reduced-order but physically interpretable representation of heat transport. Rather than recomputing the full plasma response for every magnetic configuration, one can hold $Q_{\parallel}(l,m)$ fixed as a tube-carried quantity and evaluate how the same indexed flux tubes intersect material surfaces under a modified magnetic field. This enables rapid parametric scans and provides an efficient foundation for AI training and control. In the low plasma long mean free path limit the assumed constant parallel heat flux is consistent with the so-called sheath limited regime. MPEX is no in this long mean free path regimes, so the flux tube model will be improved by adding parallel heat conduction and neutral interactions for the June demonstration.





\subsection{AI Surrogate for Helicon Hot Spot Improvement}
Generative modeling has been central to the development of modern machine learning. Unlike discriminative models, which focus on classification or label prediction, generative models learn the underlying data distribution and correlations, enabling them to produce realistic variations—such as new images, text, or scientific simulations—that resemble the training data but have not been observed before. These methods can also be adapted to domain-specific datasets (e.g., protein structures, genomic sequences, or atmospheric measurements) to model real-world phenomena and support computer-aided design. In addition, generative modeling can fuse multimodal information by learning a joint latent-space representation of heterogeneous sources, such as text, numerical simulations, 2D/3D imaging, and sequences \cite{stephens2021brief}.

\begin{figure}
    \centering
    \includegraphics[width=\linewidth]{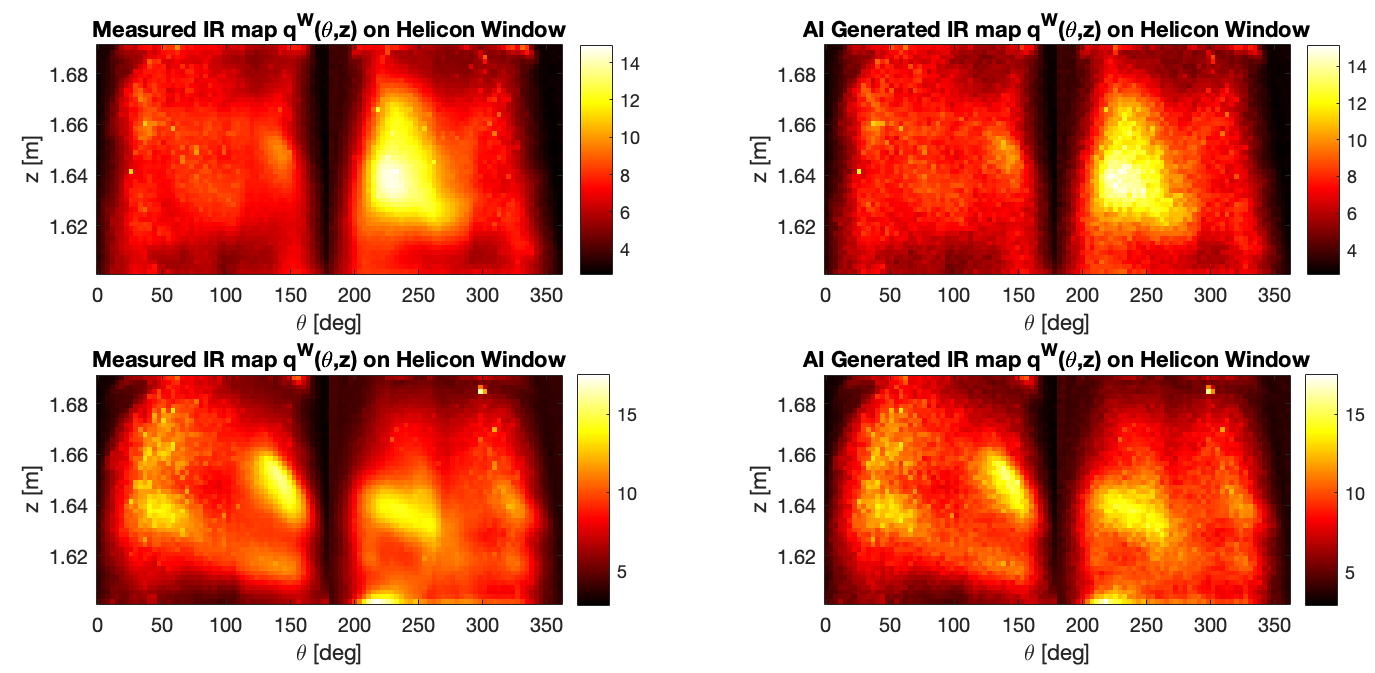}
    \caption{Comparison of IR measurements at the Helicon window with AI-generated images for the same experimental conditions. Left: two distinct Helicon measurements (top and bottom). Right: example AI samples drawn from a distribution learned to represent MPEX experimental data, corresponding to the top and bottom conditions.}
    \label{fig:VV}
\end{figure}

In this report, we use generative modeling to enable multi-objective optimization of experimental operation using multi-fidelity simulations, creating a unified probabilistic framework that bridges multiple measurement modalities with simulations at different levels of fidelity. We fuse heterogeneous data sources—experimental measurements (noisy and sparse) and multi-fidelity simulations (available at varying resolutions but potentially biased)—into a shared latent space using Variational Autoencoders (VAEs). By training on both experimental and simulated data, our models learn to represent measurement noise and aleatoric uncertainty present in laboratory experiments that idealized simulations often fail to capture, while also accounting for systematic simulation bias. We then use the resulting multi-modal models to infer control parameters, with quantified uncertainty, for MPEX operating scenarios that minimize local hot spots on the Helicon window while maximizing heat flux to the target.
\begin{figure}
    \centering
    \includegraphics[width=\linewidth]{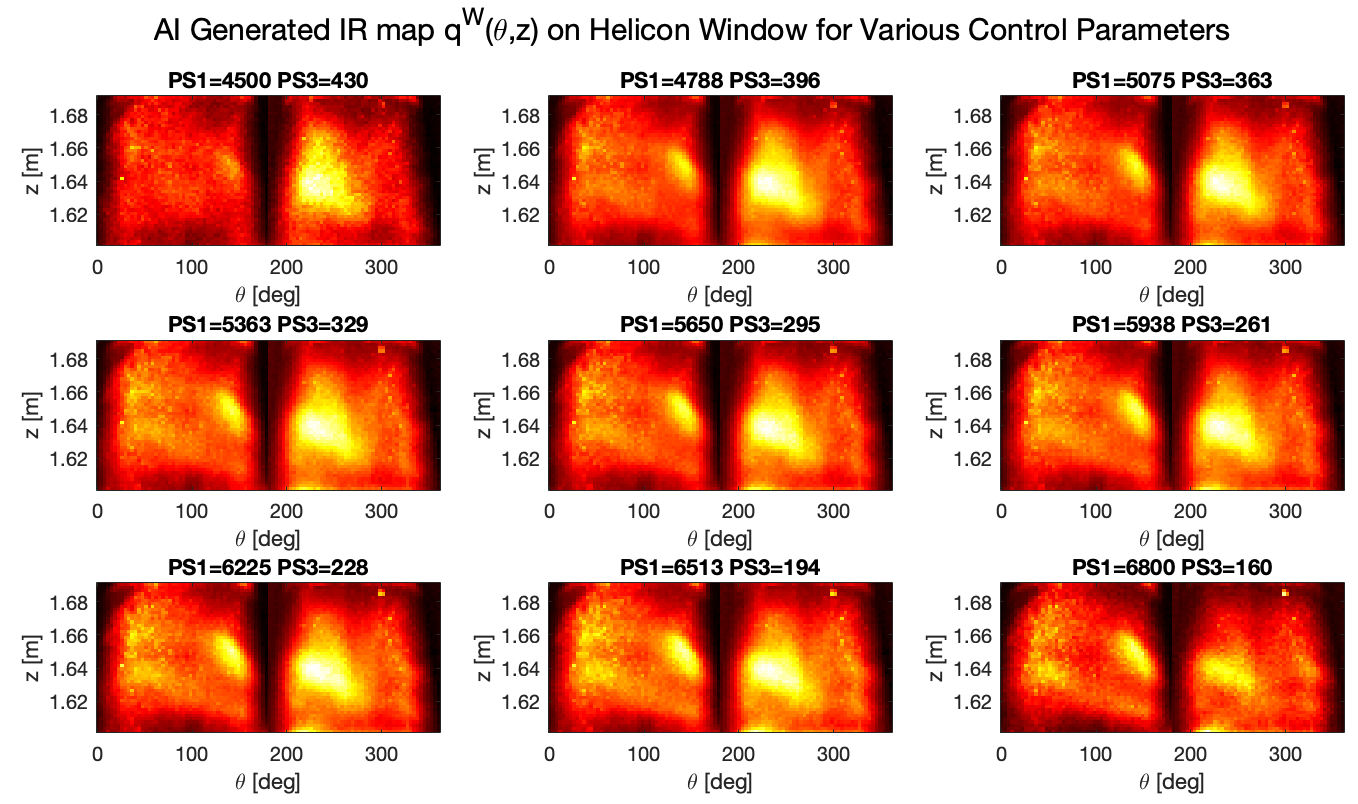}
    \caption{Depiction of a VAE used to predict experimental data that were not measured. The top-right panel shows a predicted IR map at the Helicon window for coil current settings \( \mathrm{PS1}=4500\mathrm{A} \) and \( \mathrm{PS3}=430\mathrm{A} \); this prediction is validated against experimental data in figure \ref{fig:VV}. Similarly, the bottom-left panel shows a predicted IR map for \( \mathrm{PS1}=6800\mathrm{A} \) and \( \mathrm{PS3}=160\mathrm{A} \), also validated in figure \ref{fig:VV}. The intermediate panels illustrate how the VAE interpolates to predict Helicon-window IR maps for operating conditions where observations are unavailable.}
    \label{fig:MM}
\end{figure}
\subsubsection{Generative Modeling to Explore Operation Space}
\label{subsec:GM_MPEX}
We use Variational Autoencoders (VAEs) trained on infrared (IR) measurements from Proto-MPEX. We revisit the same two IR measurements shown in Fig.~\ref{fig:proto_ir} to demonstrate the VAE’s ability to reproduce experiments performed during Proto-MPEX operation. On the left side of Fig.~\ref{fig:VV}, the two experiments are shown without the Helicon mask overlaid. The two images on the right show random samples from a VAE trained on Proto-MPEX IR data, conditioned on the corresponding device settings. The primary difference between these two experiments is the coil-current configuration: \( \mathrm{PS1}=4500\mathrm{A} \) and \( \mathrm{PS3}=430\mathrm{A} \) for the top case, and \( \mathrm{PS1}=6800\mathrm{A} \) and \( \mathrm{PS3}=160\mathrm{A} \) for the bottom case, with all other device settings unchanged.

Fig.~\ref{fig:MM} illustrates the predictive capability of the VAE for modeling experimental data. The top-left and bottom-right panels show VAE predictions for the two experiments presented in Fig.~\ref{fig:VV}. These predictions are validated against known measurements and are consistent with the same distribution as the measured data. Fig.~\ref{fig:MM} also shows how the VAE predicts unmeasured operating points for coil currents between the two experimentally observed cases. The VAE generates intermediate images by learning a smooth, continuous, and structured latent space. The latent-space regularization is enforced through the Kullback–Leibler (KL) divergence term, which encourages the latent distribution to remain close to a standard normal distribution.

\subsubsection{Approach to Multi-Modal AI Driven Optimization of MPEX Operations}
\label{ssec:mmai}
\begin{figure}
    \centering
    \includegraphics[width=\linewidth]{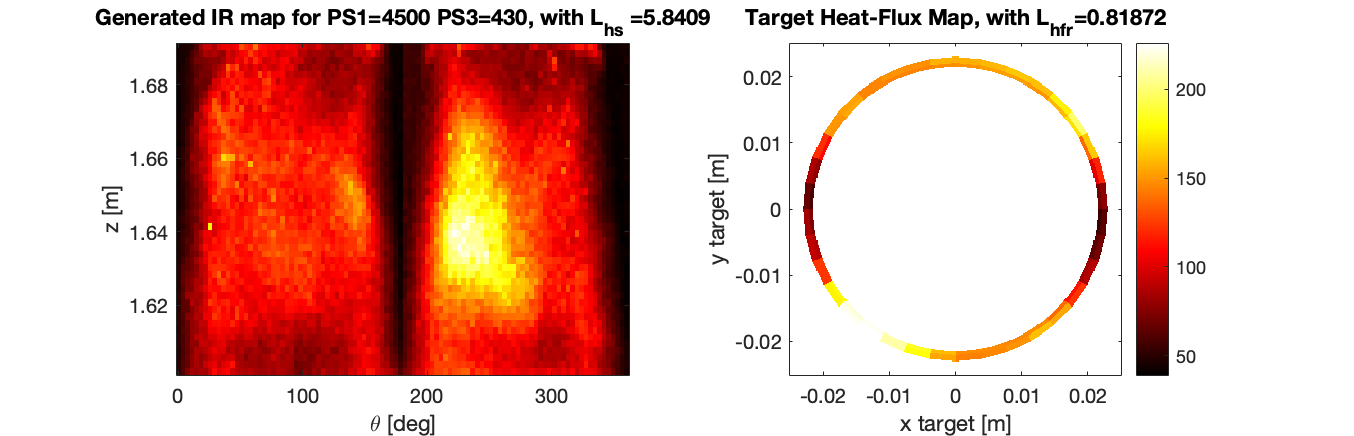}
    \caption{Left: VAE-predicted heat flux at the Helicon window for coil currents \( \mathrm{PS1}=4500\mathrm{A} \) and \( \mathrm{PS3}=430\mathrm{A} \), with the reported loss \( L_{hs} \). Right: flux-tube-resolved heat-flux mapping simulation of the target, conditioned on the Helicon-window heat flux, with the reported loss \( L_{hfr} \).}
    \label{fig:STM}
\end{figure}
We combine experimental data measured during the Proto-MPEX campaign with simulated data from the flux-tube-resolved heat-flux mapping framework described in Subsection~\ref{subsec:fluxtubemapping}. We use this multimodal information to predict optimal MPEX operating settings that minimize hot spots while maximizing the heat flux to the target.

The flux-tube-resolved heat-flux mapping framework uses the heat flux at the Helicon window to predict the heat flux at the target. We leverage the VAE described in Subsection~\ref{subsec:GM_MPEX} to predict Helicon infrared (IR) measurements from Proto-MPEX for arbitrary experimental conditions. We use the following loss function to account for hot spots at the Helicon window and the heat flux delivered to the target.

\begin{equation}
L_{hs} = \frac{\iint_{\Omega _h} \big(q^W(\theta,z) - \bar{q}^W(\theta,z)\big)^2 d\theta dz}{\bar{q}^W(\theta,z)^2} \quad \textrm{and} \quad  L_{hfr} = \frac{\iint_{\Omega _h} \big(q^W(\theta,z)d\theta dz}{\iint_{\Omega _t} \big(q^W(\theta,z)d\theta dz}.
\label{eqn:loss}
\end{equation}
Here, \( \bar{q}^{W}(\theta,z) \) denotes the average heat flux over the Helicon window, and the domains \( \Omega_h \) and \( \Omega_t \) correspond to the Helicon window and the target, respectively. Thus, \( L_{hs} \) measures the spatial variance of the heat flux on the Helicon window; minimizing \( L_{hs} \) reduces hot spots on the window. The second loss term, \( L_{hfr} \), is the ratio of the total heat flux over the Helicon window to the total heat flux at the target; minimizing \( L_{hfr} \) maximizes the heat flux delivered to the target (for a fixed window heat-flux input). 

\begin{figure}
    \centering
    \includegraphics[width=\linewidth]{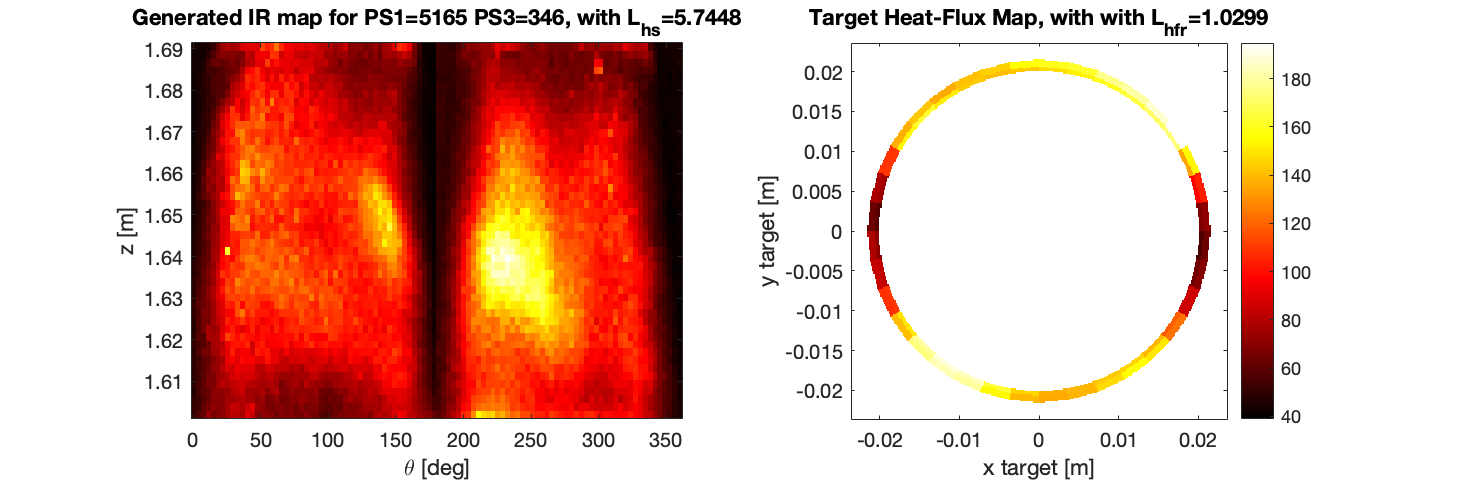}
    \caption{Left: VAE-predicted heat flux at the Helicon window for coil currents \( \mathrm{PS1}=6089\mathrm{A} \) and \( \mathrm{PS3}=254\mathrm{A} \), with the reported loss \( L_{hs} \). Right: flux-tube-resolved heat-flux mapping simulation of the target, conditioned on the Helicon-window heat flux, with the reported loss \( L_{hfr} \).}
    \label{fig:GenAI_E}
\end{figure}

For all the measure data for the proto-MPEX, figure \ref{fig:STM} depicts the minimum of $L_{hs}+L_{hfr}$. figure \ref{fig:GenAI_E} depicts an prediction, given by the multimodal framework of simulation plus generative model, where the variation of the hotspot over the helicon window is lower than seen by experiment for similar loss in heat transfer ratio.   

\clearpage
\section{MPEX AI MATERIAL ASSESSMENT DIGITAL TWIN}
\label{sec:mat}

\begin{figure}
  \centering
  \includegraphics[scale=0.4]{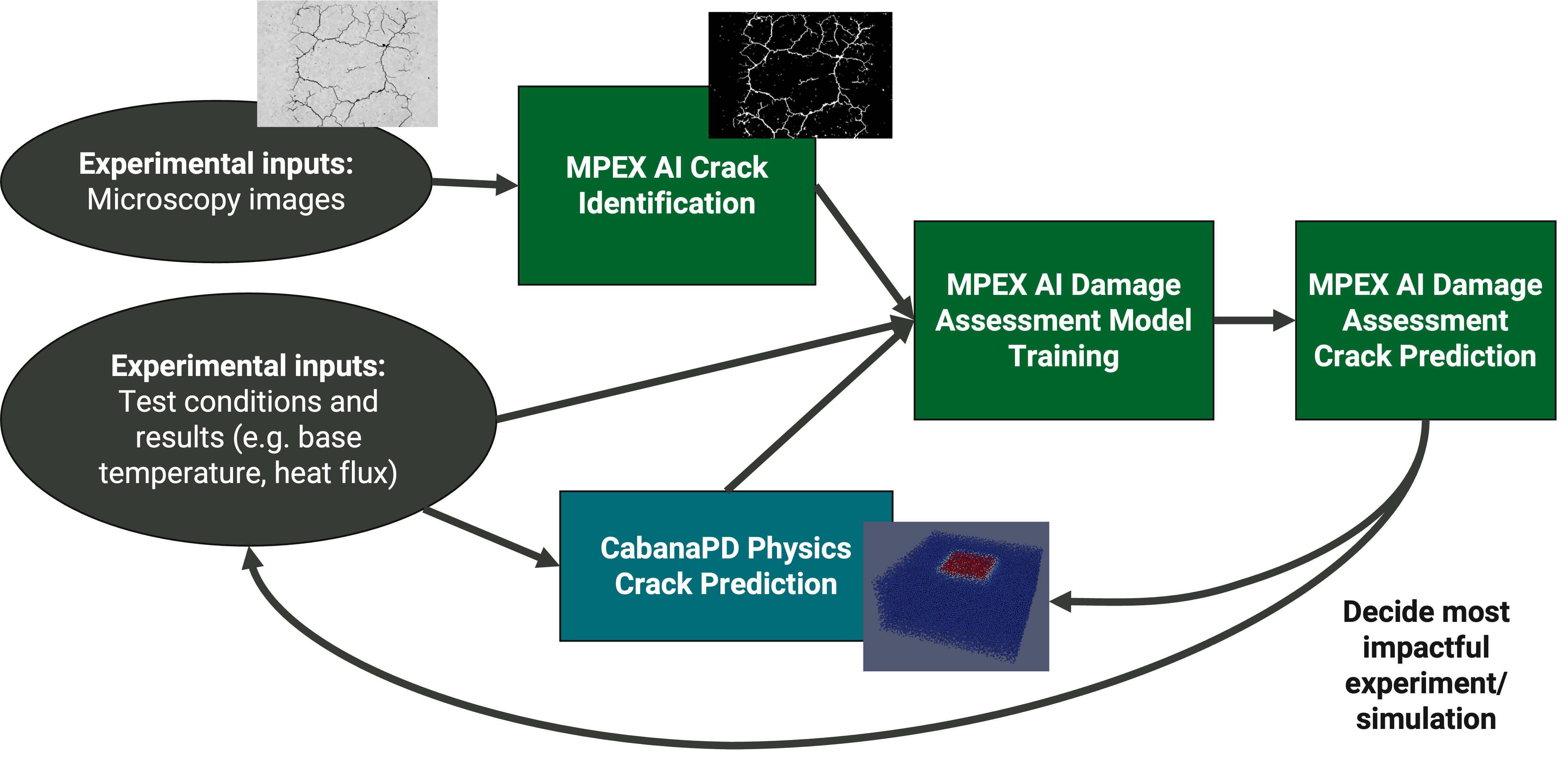}
  \caption{Workflow connections for MPEX AI Damage Assessment, including experimental inputs, physics simulation, AI prediction tools, and feedback loop for new experiments and simulations.}
\label{fig:mat:workflow}
\end{figure}

The second AI digital twin developed in this project focuses on the MPEX target materials themselves. The exposure conditions of candidate plasma facing components (PFC) are both extreme and complex, requiring a combination of experimental testing, computational modeling, and AI. We leverage throughout this effort electron beam (E-beam) experimental results as a surrogate, thermal-only test condition, in order to have reasonable performance estimates of potential MPEX material targets prior to MPEX facility operation. We thus detail the development of an E-beam AI materials surrogate and physics modeling to support it. Fig.~\ref{fig:mat:workflow} shows the workflow and connections between the AI, physics modeling, and experimental inputs for target materials. The AI development efforts span automation of the experimental data input (\textit{AI material characterization}); encoding, dimensionality reduction, and similarity distance learning of these image-based features and other experimental conditions; and forward prediction of material response to a given E-beam condition (\textit{AI material damage assessment}). Physics modeling development has focused on capability extension for material anisotropy and long-time simulation, model validation for the mechanical responses of interest for PFC materials, and simulation case preparation in order to be automatically sampled for improved model training. 

The June demonstration for the MPEX AI Damage Assessment is on track and will combine these efforts for simulation-enabled AI predictions. A sampling approach to enable learning the crack density and the boundary between regimes with and without damage, with minimal necessary simulations and experimental results is planned.

\subsection{E-beam AI Damage Assessment Digital Twin}
\label{sec:mat:ai}

We first describe the data used to develop the E-beam AI surrogate. Many scientific AI applications, and in particular materials science AI problems, struggle with data scarcity. We leverage the experimental data from E-beam experiments carried out at the JUDITH 1 facility~\cite{wirtz2013, wirtz2016, linke2011} across a range of base temperatures and heat fluxes (beam powers) on a series of tungsten alloys. Pure tungsten (``pure W'', 99.97\%), ultra-high purity tungsten (``WUHP'', 99.9999\%), potassium doped tungsten (vacuum metallizing tungsten, ``WVMW''), tungsten with 1\% tantalum (``WTa1''), and tungsten with 5\% tantalum (``WTa5'') were all tested. For each, samples were forged and cut in both the transverse and longitudinal directions, resulting in differing average grain sizes along and perpendicular to the forging direction. In addition, samples for each alloy were heat treated to recrystallize the grains for an overall uniform microstructure. Across all 15 materials (5 compositions and 3 microstructures each) scanning electron microscopy (SEM) images were taken post-exposure. 
Expert labels were provided upon viewing the SEM for cracking, surface modification, or no apparent damage.

\begin{figure}[tbp]
    \centering
    \begin{subfigure}[t]{0.45\textwidth}
        \vspace{0pt}
        \centering
        \footnotesize
        \setlength{\tabcolsep}{6pt}
        \renewcommand{\arraystretch}{1.45}

        \begin{tabular}{lcccc}
        \toprule
        & & \multicolumn{2}{c}{\# images} \\
        \cmidrule(lr){3-4}
        Material & \# experiments & total & L / R / T \\
        \midrule
        All            & 114 & 418 & 269 / 71 / 78 \\
        M192 -- W-UHP  & 32  & 100 & 77 / 11 / 12 \\
        M193 -- WVMW   & 23  & 76  & 52 / 11 / 13 \\
        M194 -- WTa1   & 17  & 67  & 39 / 12 / 16 \\
        M195 -- WTa5   & 19  & 83  & 33 / 24 / 26 \\
        M196 -- pure W & 23  & 92  & 68 / 13 / 11 \\
        \bottomrule
        \end{tabular}
        \caption{Dataset summary by material.}
        \label{fig:mat:stat}
    \end{subfigure}
    \hfill
    \begin{subfigure}[t]{0.44\textwidth}
        \vspace{0pt}
        \centering
        \includegraphics[width=\textwidth]{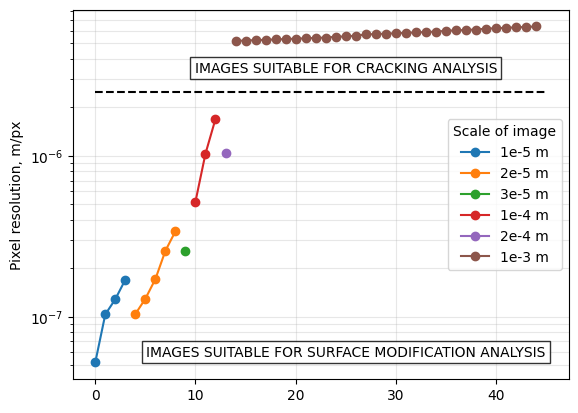}
        \caption{Pixel resolutions grouped by image scale.}
        \label{fig:mat:resolutions}
    \end{subfigure}
    \\[1.5em]
    \begin{subfigure}[t]{\textwidth}
        \vspace{0pt}
        \centering
        \includegraphics[width=0.32\textwidth]{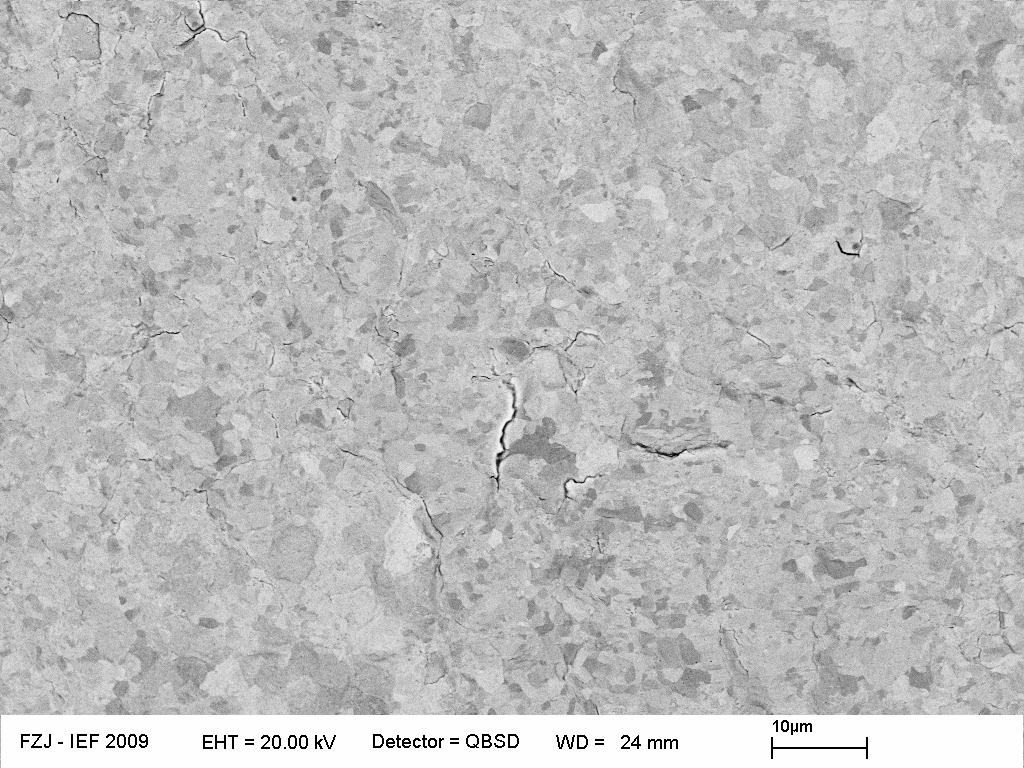}
        \hfill
        \includegraphics[width=0.32\textwidth]{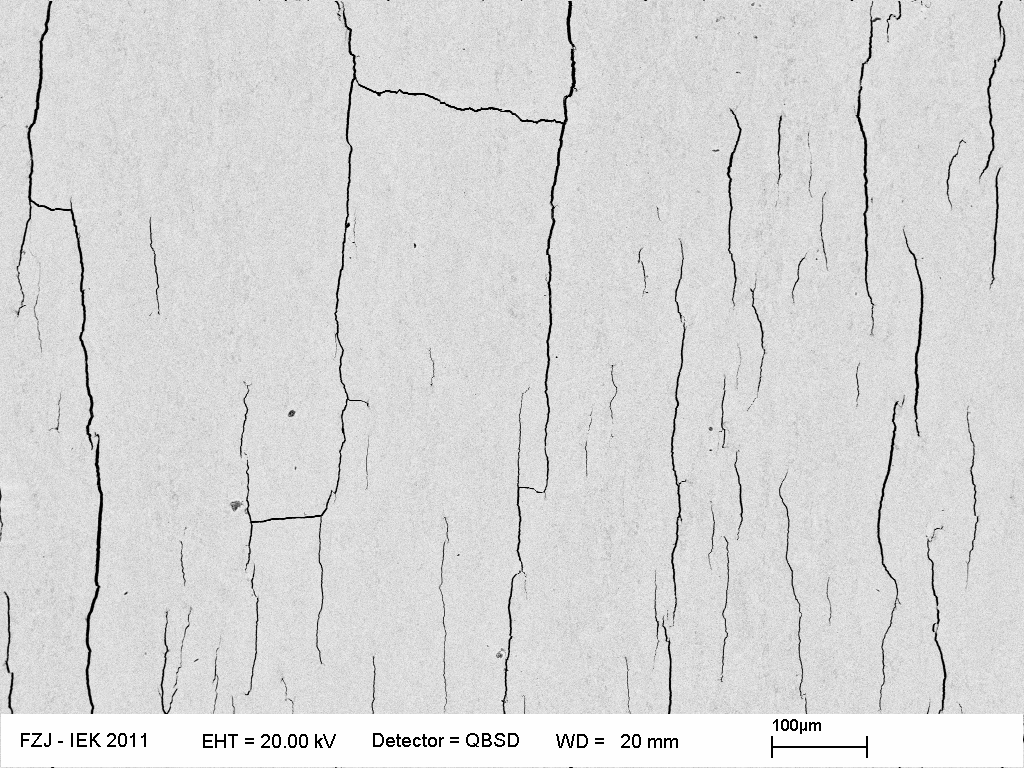}
        \hfill
        \includegraphics[width=0.32\textwidth]{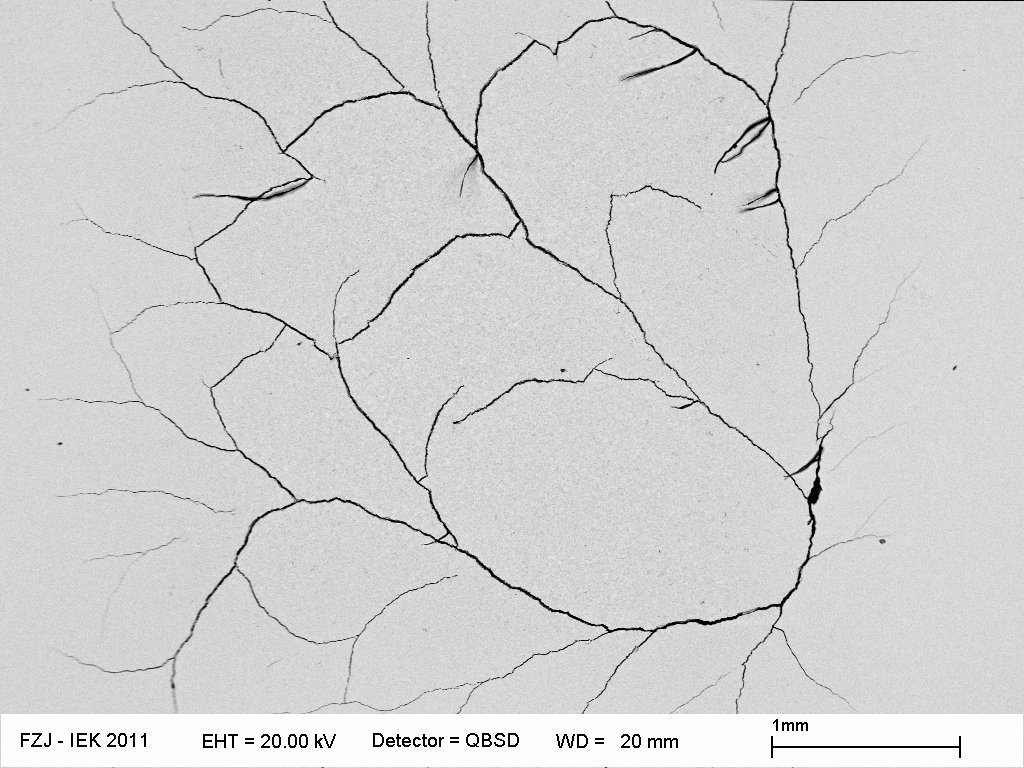}
        \caption{Representative surface scans at $10\mu m$, $100\mu m$ and $1 mm$ scales.}
        \label{fig:mat:img_examples}
    \end{subfigure}
    \\[1.5em]
    \begin{subfigure}[t]{\textwidth}
        \vspace{0pt}
        \centering
        \includegraphics[width=\textwidth]{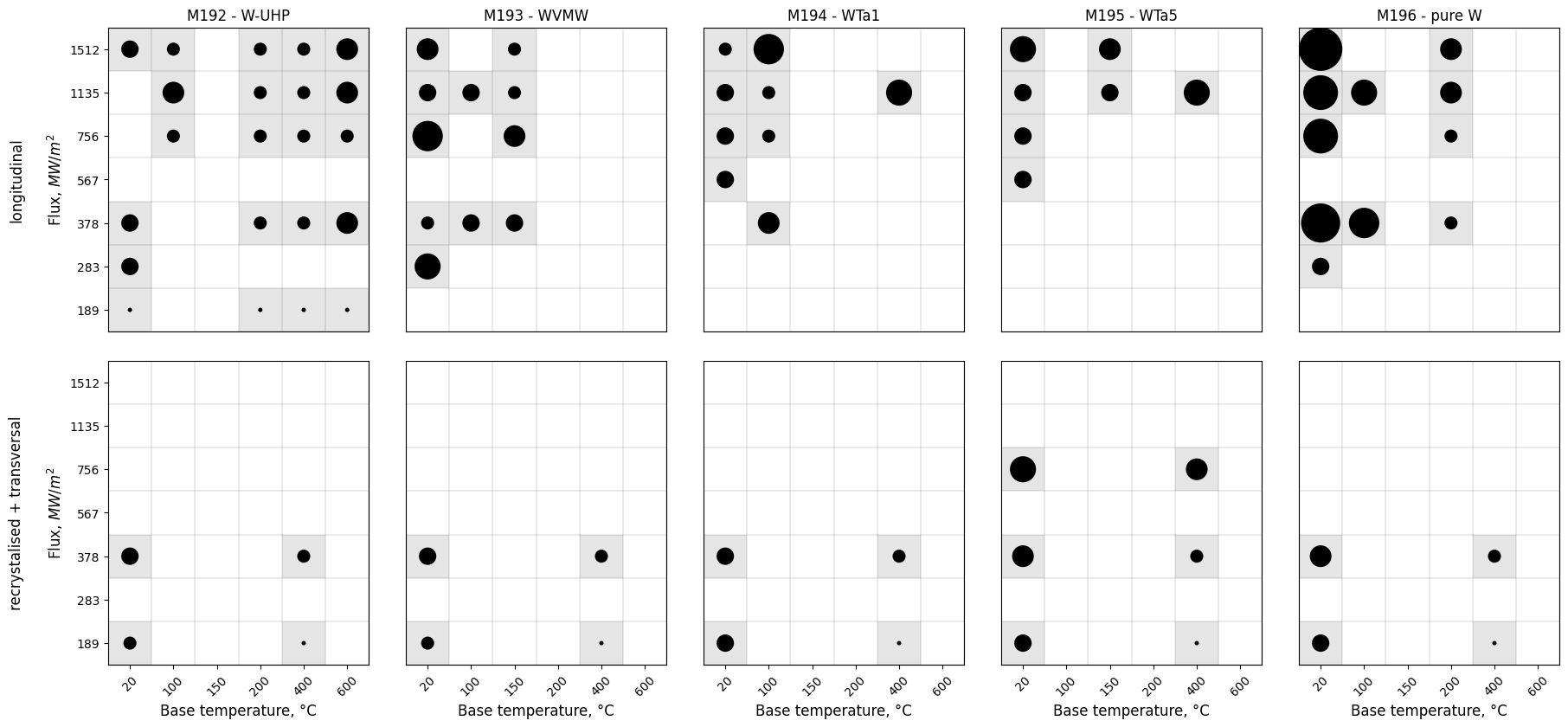}
        \caption{Test matrix of the dataset with marker size showing the relative number of collected images.}
        \label{fig:mat:test_mat}
    \end{subfigure}
    \caption{Overview of the collected surface-scan dataset. 
    (a)~Number of experiments and images per material, with image counts split into longitudinal (L), recrystallised (R), and transversal (T) cuts. 
    (b)~Pixel resolution distribution across image scales. 
    (c)~Example scans at fine and coarse scales.
    (d)~Dataset coverage across materials and exposure conditions.
    }
    \label{fig:mat:overview}
\end{figure}

An overview of the dataset is provided in Fig.~\ref{fig:mat:overview}, illustrating the distribution of experiments and images across materials, the range of image resolutions, representative surface morphologies observed after exposure, and the overall coverage of the experimental parameter space. 
Notably, the test matrix is highly sparse and unevenly sampled, with significantly fewer observations for transverse and recrystallized material samples compared to longitudinal. This experimental bias is largely purposeful: fewer tests were run on materials that failed at lower heat flux and/or base temperature.
Nonetheless this imbalance, clearly visible in the figure, poses a key challenge for data analysis and model development, as it limits the ability to learn robust, generalizable relationships across all material states and increases the risk of bias toward the more densely sampled regimes. We focus here on the development of an E-beam AI surrogate which is tolerant of these biases and of scarce data, towards the MPEX AI Damage Assessment Digital Twin.

\subsubsection{Automated Crack Identification}
\label{sec:mat:ai:cracks}

In order to leverage the full detail and richness of the experimentally characterized post-exposure fracture surfaces with AI, we require an automated image analysis framework. 
Reliable automated crack identification in SEM images is challenging due to the strong variability in image appearance across materials, imaging scales, and exposure conditions. This capability is critical for the MPEX SAS, as support for experimental analysis and decision making for MPEX candidate materials.

\begin{figure}[tb]
    \def\w{0.22\textwidth}
    \vspace{0pt}
    \centering
    \begin{tabular}{cccc}
        \textbf{\small Original} &
        \textbf{\small Sato ridge filter} &
        \textbf{\small Gaussian smoothing} &
        \textbf{\small Unsharp masking} \\[0.5em]

        \includegraphics[width=\w]{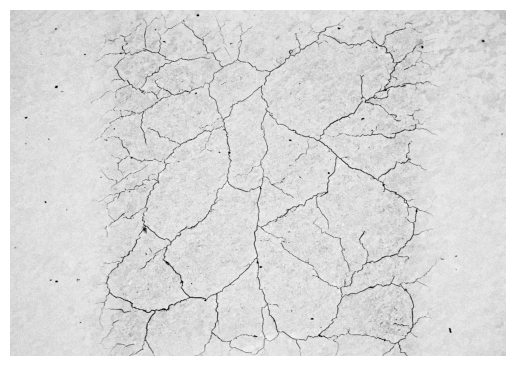} &
        \includegraphics[width=\w]{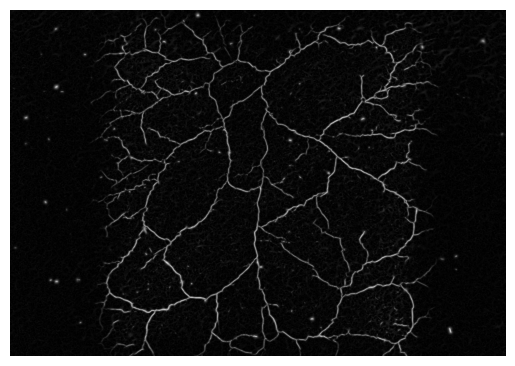} &
        \includegraphics[width=\w]{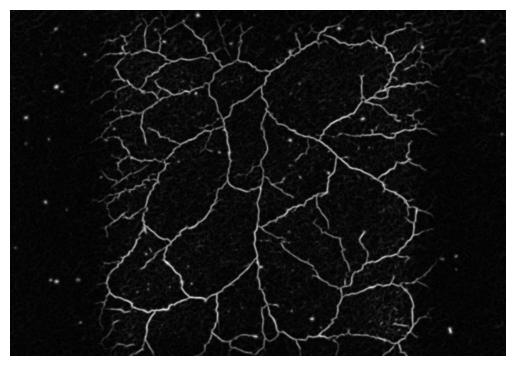} &
        \includegraphics[width=\w]{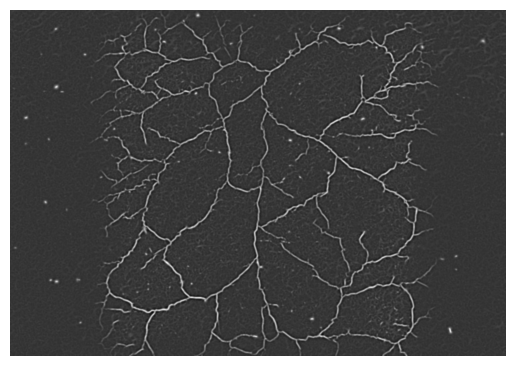} \\[1em]

        \includegraphics[width=\w]{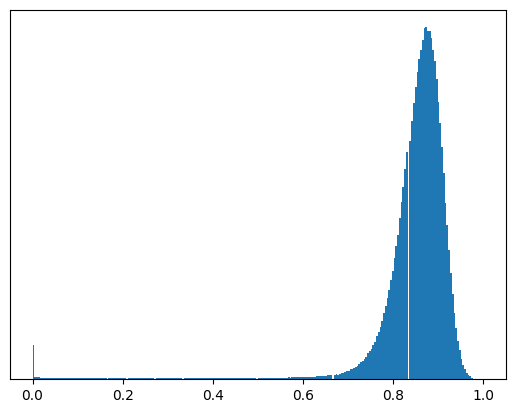} &
        \includegraphics[width=\w]{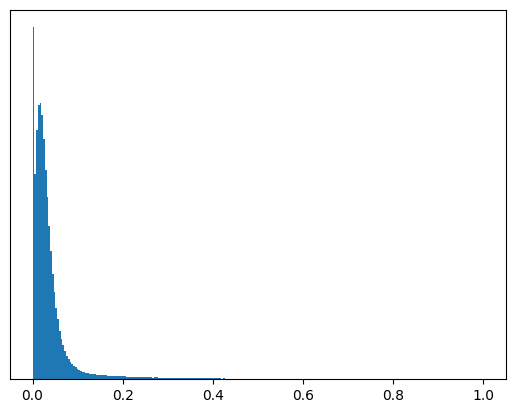} &
        \includegraphics[width=\w]{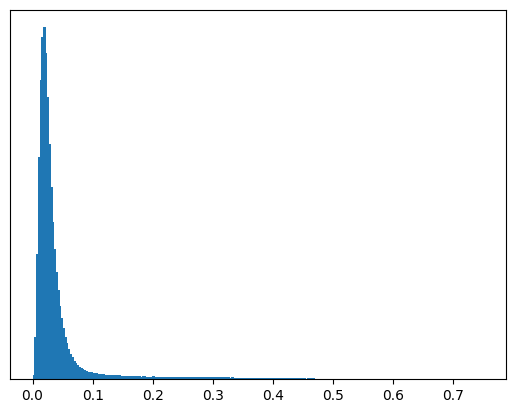} &
        \includegraphics[width=\w]{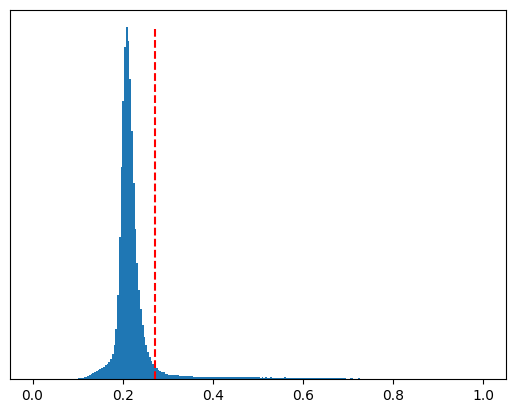}
    \end{tabular}
    \caption{Image-processing pipeline for crack detection. Top row: original image and successive enhancement steps. Bottom row: corresponding intensity histograms, showing how the distribution is progressively conditioned for robust thresholding.}
    \label{fig:mat:img_preproc}
\end{figure}

As illustrated in Fig.~\ref{fig:mat:overview}, the dataset spans a wide range of resolutions and surface morphologies, with cracks appearing at different length scales and with varying contrast relative to the background. 
Crack-like features may appear either dark or bright depending on local topography and detector response, while background texture from polishing marks, grain boundaries, or surface modification can produce intensity variations comparable to those of actual cracks. 
Consequently, simple global thresholding is insufficient: a fixed threshold may miss faint cracks in low-contrast images or over-segment textured regions in high-contrast cases.

To address these challenges, we construct a fully automated image-processing pipeline designed to robustly extract crack structures across the heterogeneous and sparsely sampled dataset. 
The pipeline aims to enhance elongated crack features while suppressing background variability, enabling consistent segmentation without image-specific parameter tuning. 
An overview of the intermediate processing steps and their effect on the image intensity distribution is shown in Fig.~\ref{fig:mat:img_preproc}.

The first stage applies \emph{Sato ridge enhancement}~\cite{SATO1998143}, which detects line-like structures by analysing second-order intensity variations via the Hessian matrix. 
Cracks, being elongated features, exhibit strong curvature perpendicular to their direction and weak variation along it. 
The filter exploits this anisotropy to enhance ridge-like responses while suppressing isotropic background structures, producing a more uniform representation of cracks across different imaging conditions.

The ridge-enhanced image is then subjected to \emph{Gaussian smoothing}. 
Beyond standard noise reduction, this step is motivated by the need to stabilize the subsequent histogram-based thresholding. 
After ridge enhancement, image histograms can become sparse or irregular, with intensities concentrated in a limited set of bins. 
Such discontinuities degrade the reliability of histogram-shape-based methods. 
Gaussian smoothing redistributes intensities through local averaging, effectively filling gaps and producing a more continuous histogram. 
As seen in Fig.~\ref{fig:mat:img_preproc}, this ``spreading'' of the histogram improves the conditioning of the segmentation problem while also suppressing high-frequency noise.

An \emph{unsharp mask} is then applied to restore local contrast attenuated by smoothing. 
This step enhances crack boundaries by amplifying local intensity differences, improving separability of crack features without reintroducing the original histogram sparsity or noise.

Segmentation is performed using \emph{triangle thresholding}~\cite{zack1977}. 
This method is well suited to the present setting, where crack pixels form a small fraction of the image and appear as a low-population tail extending from a dominant background mode. 
In such unimodal but skewed histograms, variance-based methods such as Otsu’s method can become unreliable, particularly when the foreground is sparse~\cite{otsu1979, sezgin2004}. 
The triangle method instead determines the threshold geometrically from the histogram shape, providing a robust and fully automatic separation of crack and background regions.

Post-processing is applied to refine the resulting binary mask. 
Small isolated components are removed to eliminate spurious detections arising from residual texture or noise. 
Subsequently, \emph{morphological closing} connects nearby crack fragments and fills small gaps, producing more continuous crack networks. 
For certain analyses, an optional \emph{skeletonization} step reduces the crack regions to one-pixel-wide representations while preserving topology, enabling characterization of crack connectivity and length independently of width.

\begin{figure}[tb]
  \centering
  \includegraphics[width=0.8\textwidth]{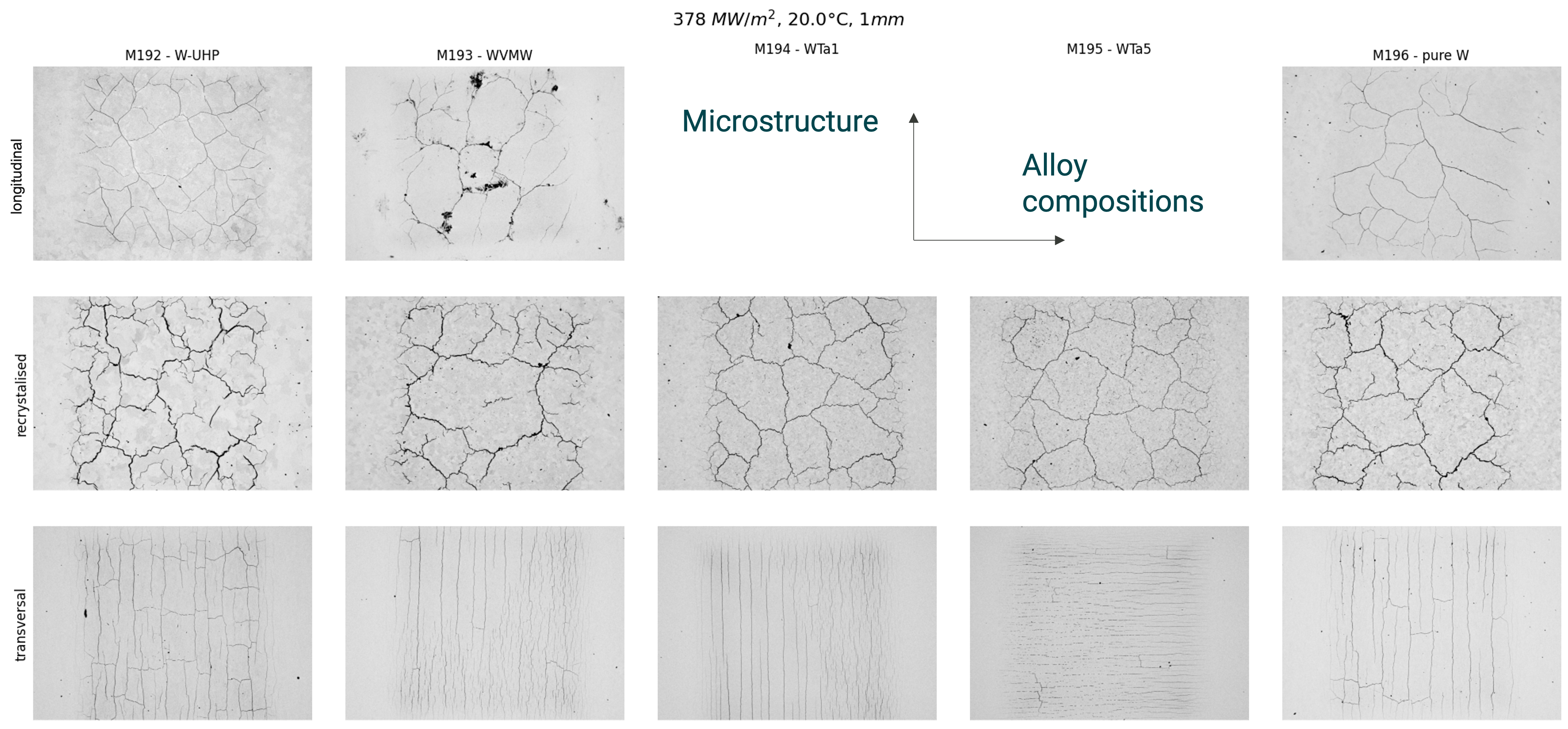}
  \rule{0.8\textwidth}{0.5pt}
  \\[1em]
  \includegraphics[width=0.8\textwidth, trim=0 0 0 1cm, clip]{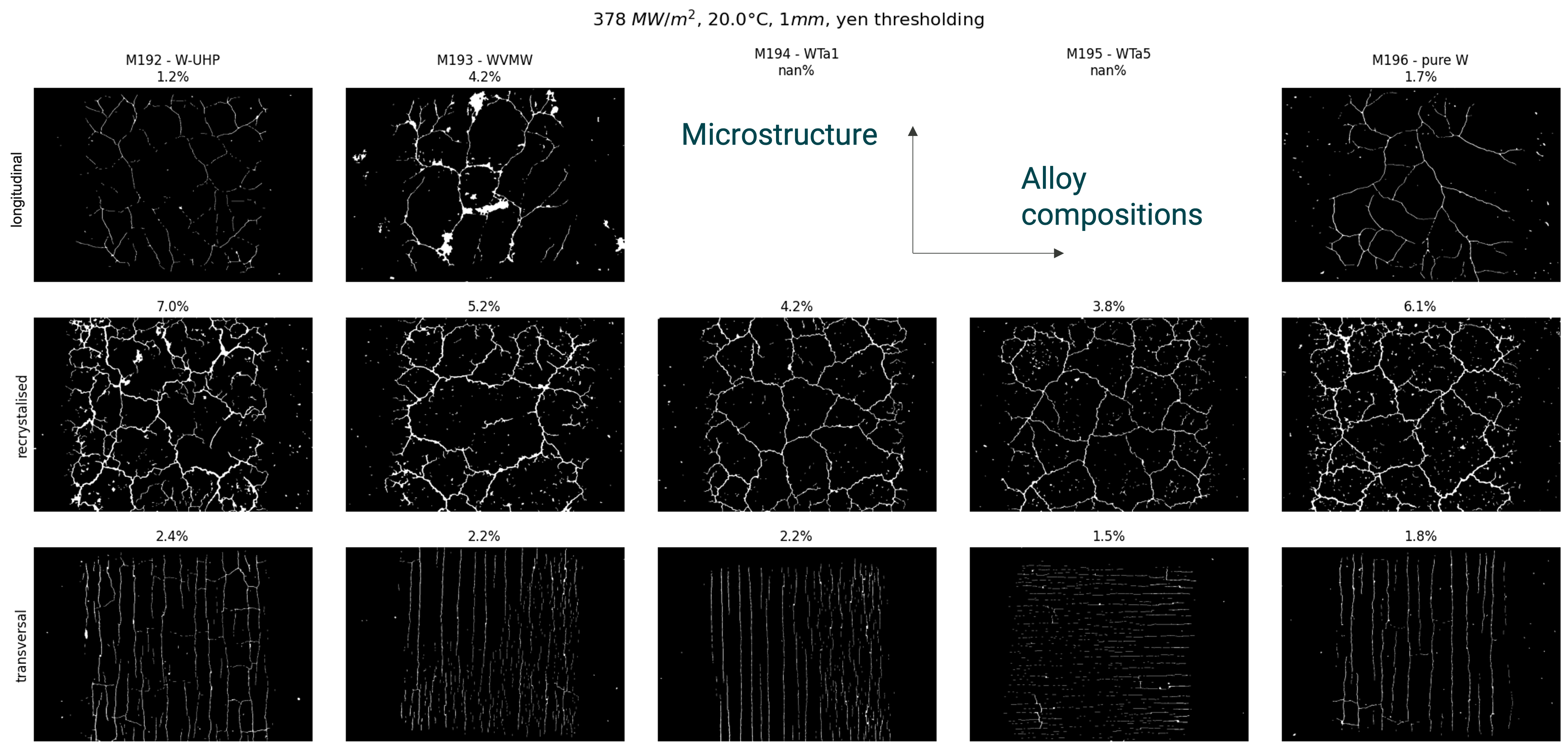}
  \caption{Top row: Experimental characterization for a single heat flux and base temperature condition showing significant cracking. One representative SEM image shown per microstructure and alloy. Bottom row: Automated crack extraction for use within the MPEX AI Damage Assessment surrogate.}
\label{fig:mat:sem}
\end{figure}

Finally, the processed mask is used to compute a \emph{crack density} metric, defined as the fraction of the image area identified as crack. 
This provides a consistent scalar measure of damage severity that can be compared across materials and exposure conditions.
Examples of the resulting crack extraction are shown in Fig.~\ref{fig:mat:sem}, demonstrating the ability of the pipeline to recover crack networks across different alloys and microstructures despite substantial variability in image appearance.

Overall, the proposed pipeline, combining ridge enhancement, histogram conditioning, adaptive thresholding, and morphological refinement, enables robust and fully automated crack identification across a heterogeneous and sparsely sampled dataset without manual tuning. The initial deployment of this tool is described in Section \ref{sec:workflows} and will be demonstrated for future use within the MPEX SAS. 

\subsubsection{Crack Prediction}
\label{sec:mat:damage}
We next detail the steps taken to generate an AI framework for damage prediction and assessment that is robust across available inputs and extensible to additional inputs (e.g., simulation results).
This begins with heterogeneous SEM data, requiring a representation that is consistent across imaging conditions and resolutions, and that captures the underlying morphology of surface features.
The key idea is to map images to a feature space in which similarity reflects physical damage characteristics, and then relate this space back to experimental conditions.

\begin{figure}[tb]
    \vspace{0pt}
    \centering
    \includegraphics[width=\textwidth]{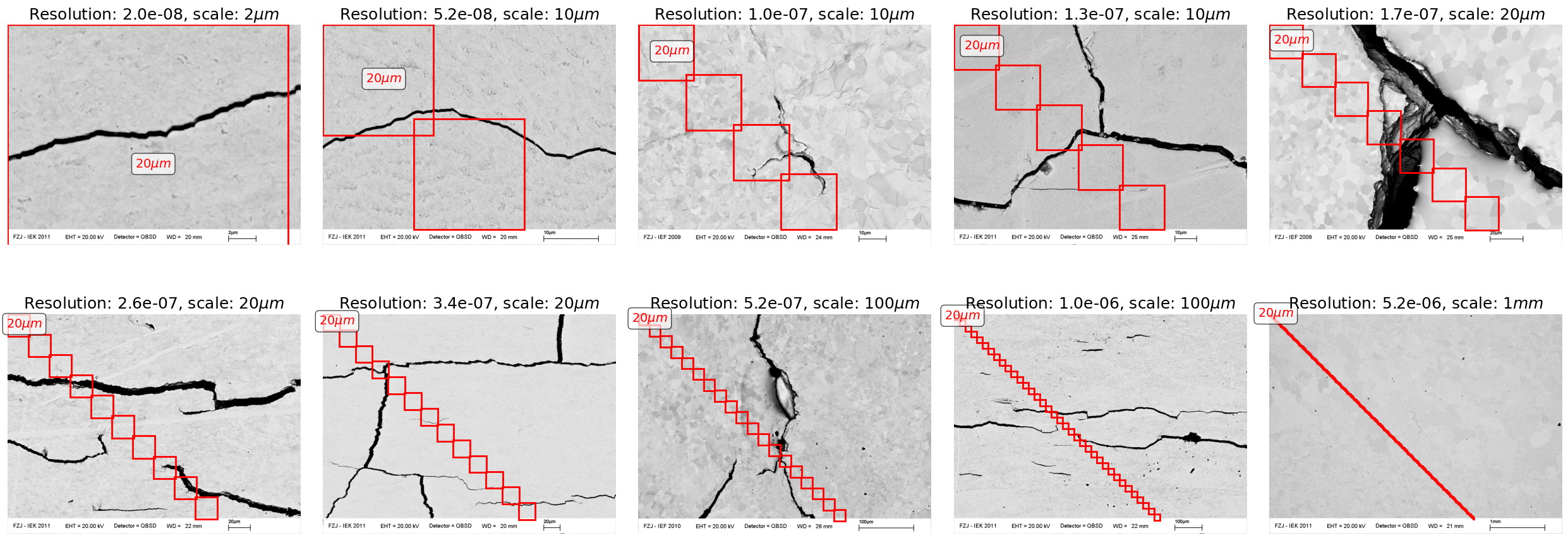}
    \rule{\textwidth}{0.5pt}
    \\[1em]
    \includegraphics[width=\textwidth]{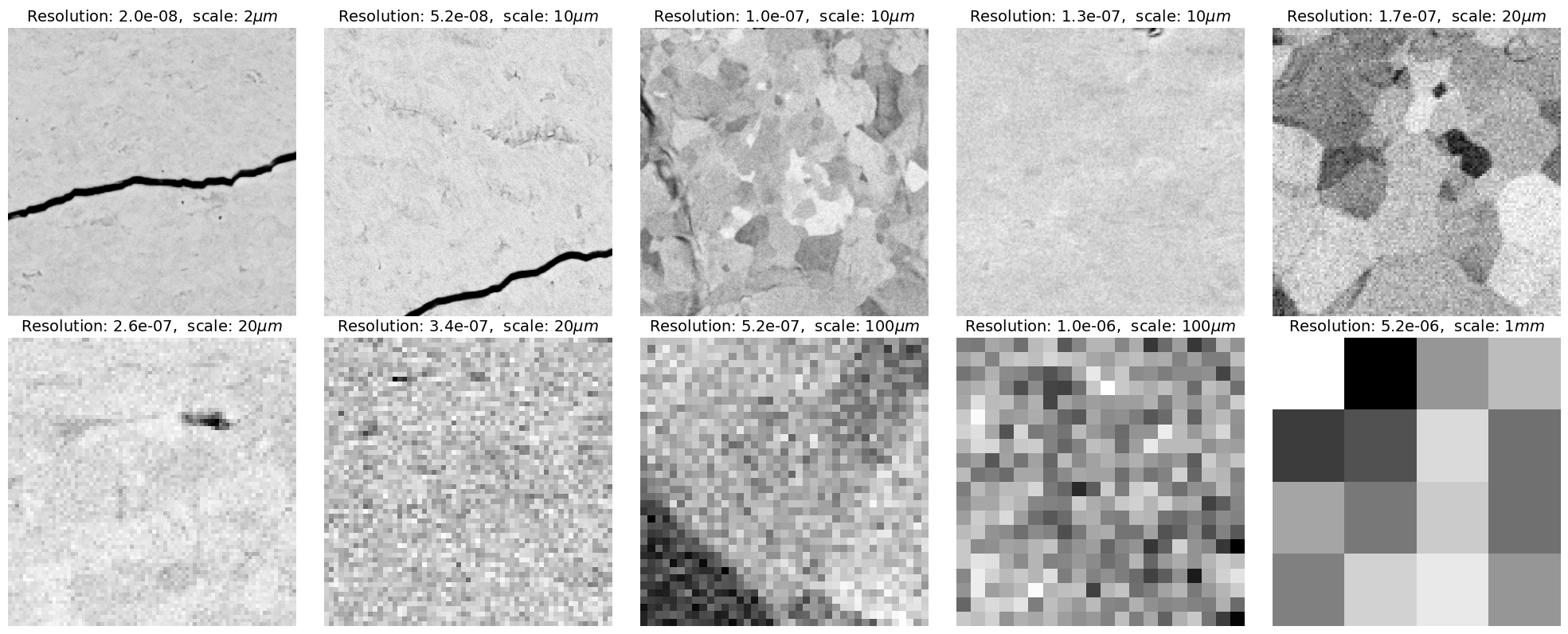}
    \caption{Top row: Representative SEM images at different resolutions with overlaid patch boundaries corresponding to a fixed physical field of view ($20\,\mu m$), ensuring consistent sampling across resolutions. 
    Bottom row: Representative patches at different resolutions.  Low-resolution patches are under-resolved and do not reliably capture surface morphology, and are therefore excluded from surface analysis.}
    \label{fig:mat:patches}
\end{figure}

\paragraph{Patch-based representation.}
Given the wide range of image resolutions present in the dataset (Fig.~\ref{fig:mat:overview}), direct processing of full images is not appropriate, as the physical scale represented by a fixed number of pixels varies significantly across samples. 
Instead, images are decomposed into patches defined by a fixed \emph{physical field of view}, typically $50\,\mu\mathrm{m}$ or $100\,\mu\mathrm{m}$, as illustrated in Fig.~\ref{fig:mat:patches}.
The corresponding patch size in pixels therefore varies depending on the image resolution.


This choice serves two complementary purposes.
First, it ensures that each patch represents a comparable physical region of the material surface, regardless of the original image resolution. 
As a result, morphological features such as crack density, crack width, and connectivity are represented consistently across samples.
This is essential in a multi-resolution dataset: patching in pixel space would otherwise mix fundamentally different physical scales, leading to representations that are not directly comparable.

Second, the use of physically defined patches aligns with the characteristic length scales at which surface damage and texture manifest.
Crack networks, grain-dependent surface modification, and other morphological features typically exhibit spatial organization over finite, physically meaningful regions rather than across the entire field of view.
By selecting patch sizes on the order of tens of microns, we capture these representative local patterns while avoiding dilution by larger-scale heterogeneity or boundary effects.
In this sense, each patch can be viewed as a localized ``sample'' of the surface state, containing sufficient contextual information to characterize the underlying damage mechanisms.

Together, these considerations motivate a patch-based representation that is both scale-consistent across images and well matched to the intrinsic spatial structure of the observed surface damage.

\paragraph{Feature encoding using vision transformers}

Each image patch is encoded into a high-dimensional feature vector using a pre-trained DINOv2 vision transformer model~\cite{Oquab_2024_TMLR_DINOv2, Caron_2021_ICCV}. 
DINOv2 is a self-supervised foundation model trained on large-scale image data to produce general-purpose visual representations without requiring task-specific supervision. 
Such models capture both local texture and global structural information, making them well suited for representing complex surface morphologies in SEM images, including crack networks and surface modification patterns.

A key advantage of DINOv2 is its ability to generalize across domains. 
Although trained on natural images, DINOv2 features have been shown to transfer effectively to scientific imaging tasks, where labeled data are scarce and image statistics differ significantly from standard benchmarks. 
In materials science, pre-trained vision transformers, including DINOv2, have been successfully used to represent microstructures and learn microstructure--property relationships without task-specific retraining~\cite{Whitman_2024_CVPR, WhitmanLatypov_2025_ActaMaterialia}. 
Similarly, recent work demonstrates that DINOv2-based representations are effective for microscopy and biological imaging, capturing morphology-driven patterns and enabling analysis in low-label regimes~\cite{Moutakanni_2025_CellDINO_PLOSCompBio}. 
Applications to geoscience and medical imaging further show that DINOv2 provides robust representations under strong domain shift and heterogeneous imaging conditions~\cite{BrondoloBeaussant_2025_JRMGE, MullerFranzes_2025_SciRep_MST}.

In addition, DINOv2 produces meaningful \emph{patch-level} embeddings, which are particularly relevant for the present work. 
Crack structures are inherently local and spatially organized, and similarity between patches is more informative than global image-level similarity. 
Recent studies on anomaly detection and scientific imaging have shown that distances between DINOv2 patch embeddings provide a powerful signal for identifying structural differences and defects~\cite{Damm_2025_WACV}. 
This makes DINOv2 a natural choice for encoding SEM patches, where the goal is to compare local surface morphology across materials and exposure conditions.

Overall, the combination of strong transferability, robustness to data scarcity, and the ability to capture fine-scale morphological structure motivates the use of DINOv2 as a feature encoder for SEM-based damage analysis.

\paragraph{Dimensionality reduction and manifold representation.}

\begin{figure}[tb]
    \vspace{0pt}
    \centering
    \includegraphics[width=\textwidth, trim=0 0 16.5cm 0, clip]{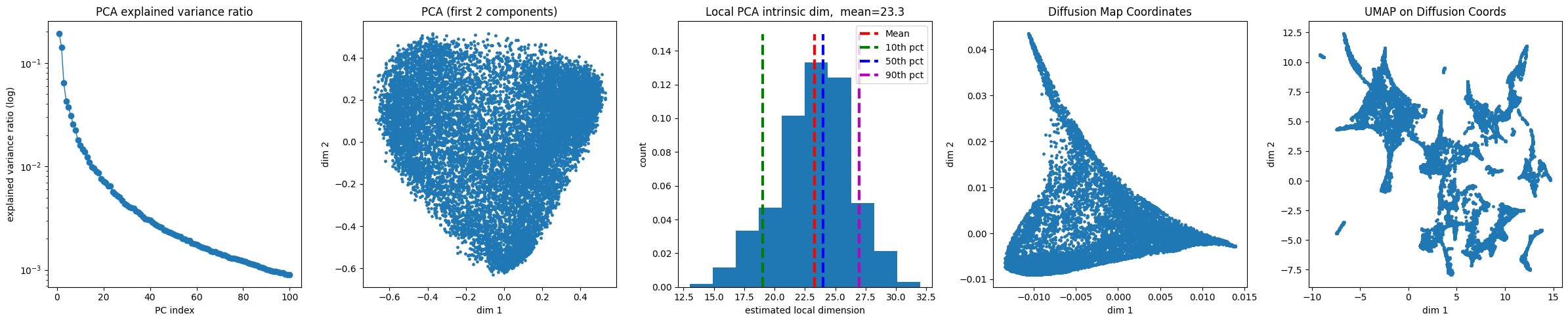}
    \caption{Embedding pipeline for DINOv2 patch features. 
    From left to right: PCA variance decay, PCA projection, intrinsic dimension estimates, and diffusion map coordinates.}
    \label{fig:mat:embeddings}
\end{figure}

The DINOv2 embeddings are high-dimensional and contain significant redundancy, motivating dimensionality reduction prior to downstream analysis (Fig.~\ref{fig:mat:embeddings}). 
We first apply principal component analysis (PCA) and retain the number of components required to explain $95\%$ of the variance~\cite{Jolliffe2002PCA}. 
This step reduces noise and compresses the representation while preserving the dominant modes of variation in the data.

However, PCA is inherently a linear method and may not capture the intrinsic structure of complex image representations. 
In particular, features derived from natural images and scientific imaging data often lie on a low-dimensional nonlinear manifold embedded in a high-dimensional space. 
To capture this structure, we further transform the PCA-reduced features using \emph{diffusion coordinates}~\cite{Coifman2006DiffusionMaps, Coifman2005GeometricDiffusions}. 
Diffusion coordinates construct a low-dimensional embedding by modeling a diffusion process on a graph defined by pairwise similarities between data points. 
Distances in this space correspond to \emph{diffusion distances}, which reflect connectivity along the data manifold rather than simple Euclidean proximity.

To determine the appropriate dimensionality of the diffusion embedding, we estimate the intrinsic dimension of the data using local PCA. 
This provides a data-driven estimate of the number of degrees of freedom required to represent the underlying structure, ensuring that the diffusion map captures the essential geometry without introducing unnecessary dimensions.

This approach is particularly well suited to DINOv2 features. 
Although high-dimensional, these embeddings exhibit strong geometric organization, with samples arranged according to semantic and morphological similarity. 
Diffusion maps exploit this structure by preserving relationships along the manifold, allowing patches with similar surface morphology, such as crack patterns or texture features, to be mapped close to each other, even if their raw feature vectors differ significantly.

From a broader perspective, this procedure can be interpreted as a form of \emph{manifold learning}, where the goal is to uncover a low-dimensional representation that captures the intrinsic geometry of the data~\cite{Tenenbaum2000Isomap, Belkin2003LaplacianEigenmaps, Roweis2000LLE}. 
Unlike purely linear methods, nonlinear manifold learning techniques are able to represent complex variations in data that arise from underlying physical processes, such as crack formation and surface evolution. 
This is particularly important in the present setting, where the relationship between observed morphology and underlying material or exposure conditions is highly nonlinear.


In the resulting diffusion space, Euclidean distances provide a meaningful measure of similarity between image patches in terms of their morphology. 
This representation forms the basis for subsequent similarity-based learning and enables robust comparison of samples across different materials, resolutions, and experimental conditions.

\paragraph{Metric learning in parameter space.}

\begin{figure}[tb]
    \vspace{0pt}
    \centering
    \includegraphics[width=0.32\textwidth]{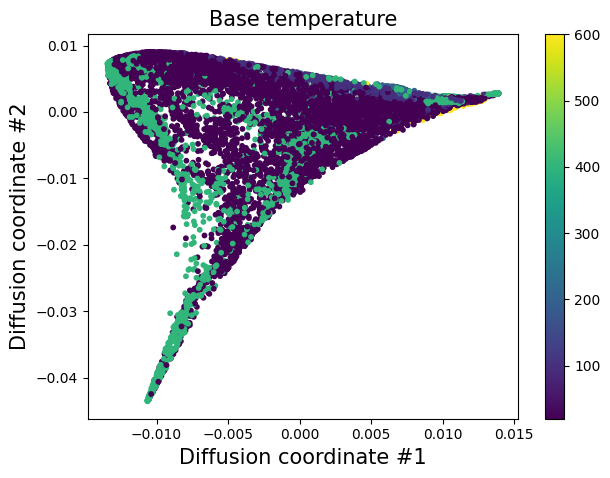}
    \hfill
    \includegraphics[width=0.32\textwidth]{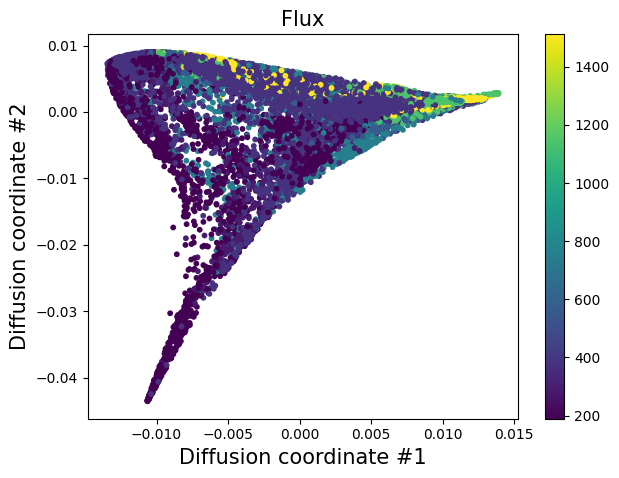}
    \hfill
    \includegraphics[width=0.32\textwidth]{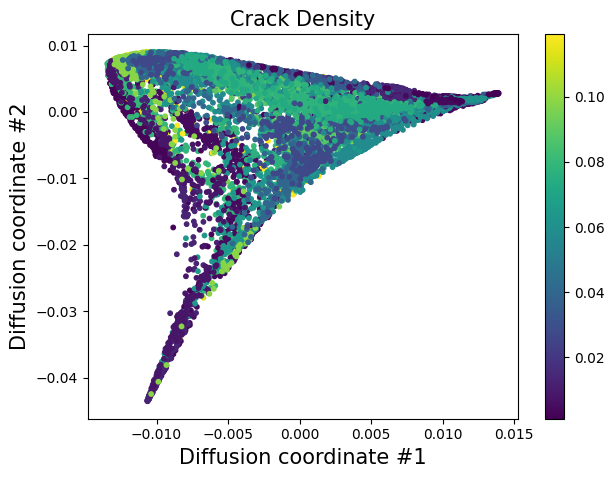}
    \caption{Patch embeddings in diffusion coordinate space colored by base temperature, heat flux, and crack density, which exhibit clear spatial correlation in the embedding.}
    \label{fig:mat:inout_correlation}
\end{figure}

Once patches are embedded in diffusion space, distances between patches provide a natural measure of similarity in terms of observed surface morphology. 
This is fundamentally different from distances in the original material-property or exposure-parameter space, where similarity is defined in terms of experimental conditions (e.g., base temperature, heat flux, alloy composition). 
As illustrated in Fig.~\ref{fig:mat:inout_correlation}, proximity in the embedding space correlates with both experimental conditions and crack density, indicating that the learned representation organizes patches according to physically meaningful structure.

In practice, morphological similarity does not necessarily correlate directly with proximity in parameter space, particularly in sparse and nonlinear regimes such as those shown in Fig.~\ref{fig:mat:overview}. 
Therefore, learning directly in parameter space may fail to capture relevant relationships between samples. 
By contrast, the feature space encodes the observed outcome (damage morphology), making it a more appropriate space for defining similarity between samples.

While similarity is most naturally defined in feature space, the ultimate goal is to construct a predictive model from experimental parameters to quantities of interest such as crack density. 
To bridge this gap, we learn a metric in parameter space that aligns with distances in feature space.

Let $p \in \mathbb{R}^d$ denote the vector of material and exposure parameters, and let $z \in \mathbb{R}^k$ denote the corresponding embedding in diffusion space.
We define a parameter-space distance
\begin{align*}
    d^2_C(p_i, p_j) = (p_i - p_j)^T C^{-1} (p_i - p_j),
\end{align*}
where $C$ is a positive definite matrix to be learned. 
The goal is to choose $C$ such that distances in parameter space approximate distances in feature space, i.e.,
\begin{align*}
    d(p_i, p_j) \approx \|z_i - z_j\|.
\end{align*}
This can be formulated as an optimization problem over $C$:
\begin{align*}
    \min_{C \succ 0} \sum_{i,j} \left( d^2_C(p_i, p_j) - \|z_i - z_j\|^2 \right)^2.
\end{align*}
This approach can be interpreted as a form of metric learning, where the goal is to embed the parameter space in a way that reflects the geometry of the feature space.
In effect, we learn which directions in parameter space are most relevant for predicting damage morphology.

\paragraph{Kernel-based prediction in learned metric space.}

Once the metric in parameter space has been learned, it induces a corresponding similarity kernel that can be used for prediction. 
Given two parameter vectors $p_i$ and $p_j$, we define a radial kernel of the form
\begin{align*}
    K(p_i, p_j) = \exp\left(-\frac{(p_i - p_j)^T C^{-1} (p_i - p_j)}{\sigma^2}\right),
\end{align*}
where $C$ is the learned positive definite matrix and $\sigma$ is a bandwidth parameter. 
This kernel encodes similarity between samples in a way that is consistent with distances in the diffusion feature space.

A simple approach to prediction is the Nadaraya--Watson kernel smoother,
\begin{align*}
    \hat{y}(p) = \frac{\sum_i K(p, p_i)\, y_i}{\sum_i K(p, p_i)},
\end{align*}
which provides a local, weighted average of observed values~\cite{Nadaraya1964, Watson1964}. 
While effective, this estimator can be sensitive to noise and does not explicitly regularize the function being learned.

A more general and robust formulation is \emph{kernel ridge regression} (KRR), which seeks a function of the form
\begin{align*}
    \hat{y}(p) = \sum_i \alpha_i K(p, p_i),
\end{align*}
where the coefficients $\alpha_i$ are obtained by solving
\begin{align*}
    (K + \lambda I)\alpha = y.
\end{align*}
Here, $K$ is the kernel matrix with entries $K_{ij} = K(p_i, p_j)$ and $\lambda > 0$ is a regularization parameter. 
Kernel ridge regression can be interpreted as a regularized least-squares problem in a reproducing kernel Hilbert space, providing improved stability and generalization compared to simple kernel smoothing, particularly in sparse or noisy regimes~\cite{Saunders1998, Murphy2012,Scholkopf2002LearningWithKernels}.

\begin{figure}[tb]
    \vspace{0pt}
    \centering
    \begin{tabular}{ccc}
        \textbf{\small Diffusion Features} &
        \textbf{\small Raw Inputs} &
        \textbf{\small Learned Metric} \\[0.5em]
        \includegraphics[width=0.32\textwidth]{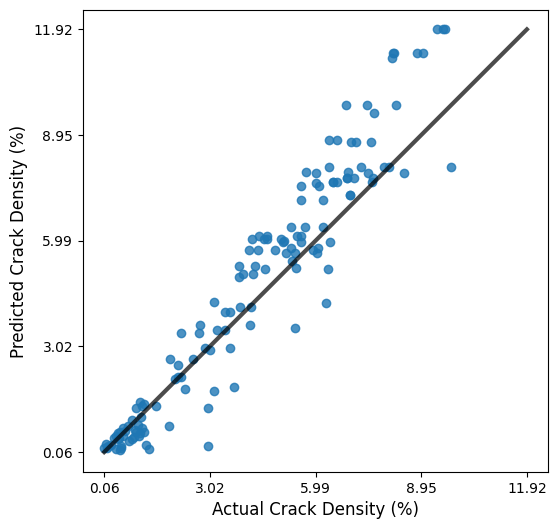}&
        \includegraphics[width=0.32\textwidth]{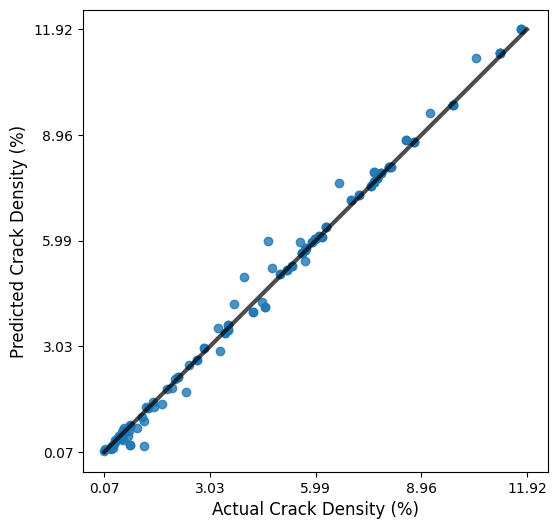}&
        \includegraphics[width=0.32\textwidth]{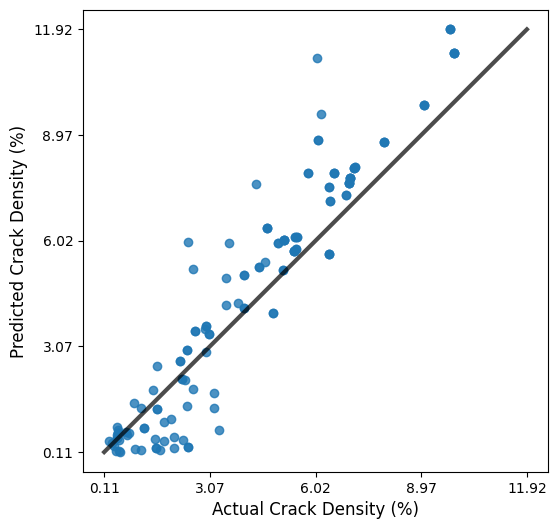}
    \end{tabular}
    \caption{Parity plots of predicted versus observed crack density for kernel ridge regression using diffusion features, raw input parameters, and input parameters with a learned metric.}
    \label{fig:mat:parities}
\end{figure}

The effect of different similarity definitions on predictive performance is illustrated in Fig.~\ref{fig:mat:parities}. 
Regression directly on input parameters yields near-perfect parity; however, this behavior reflects overfitting to the limited and discrete parameter combinations rather than capturing morphology-driven relationships. 
In contrast, models based on diffusion features and the learned metric produce more consistent and physically meaningful predictions, with the learned metric closely matching feature-space performance. 
A small bias remains, with the learned-metric model slightly underestimating crack density in the low-damage regime.

\begin{figure}[tb]
  \centering
  \includegraphics[width=0.8\textwidth]{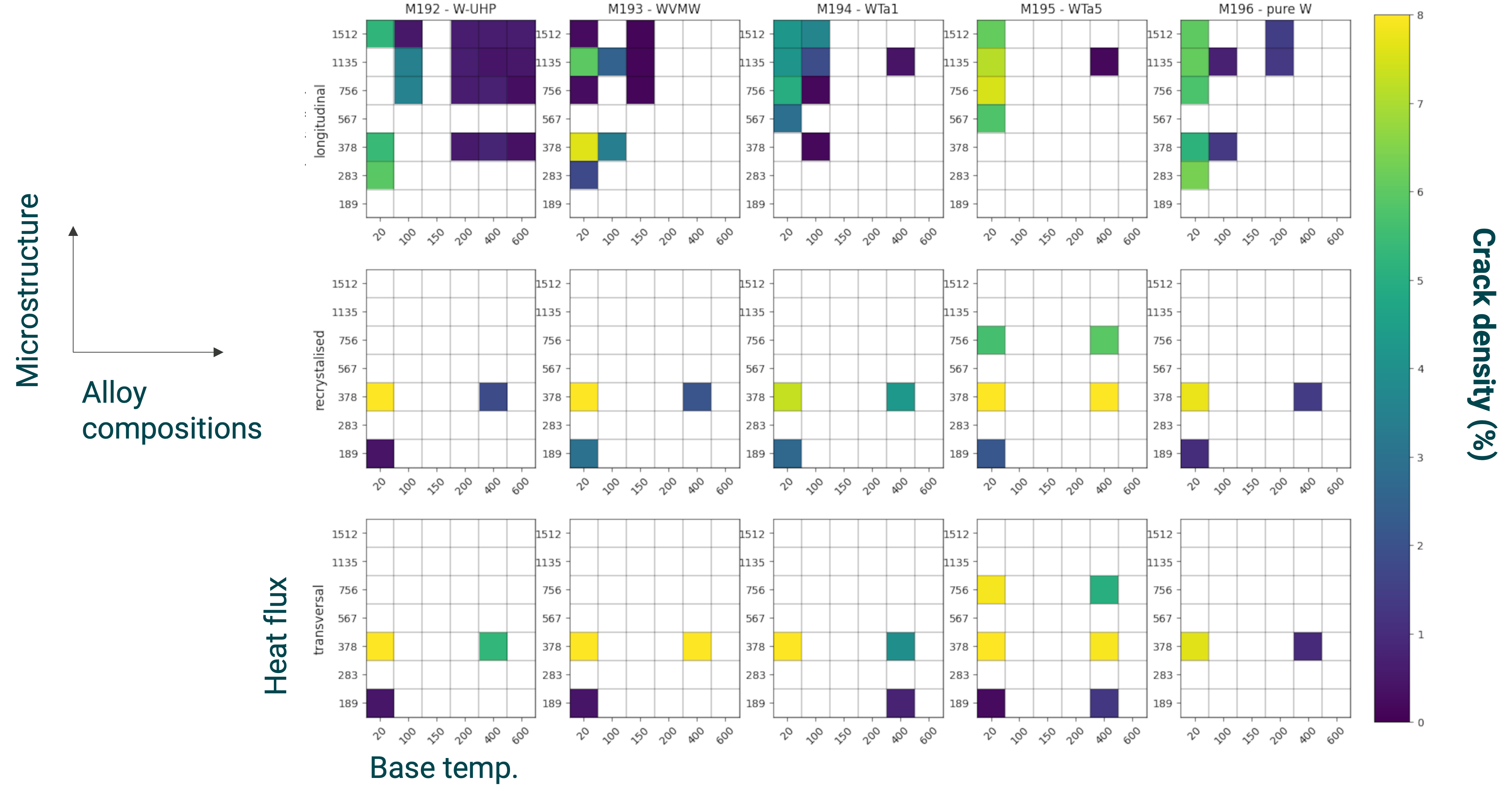}
  \rule{\textwidth}{0.5pt}
  \\[0.2em]
    \begin{tabular}{ccc}
        \textbf{\scriptsize Regression with raw inputs} &
        \textbf{\scriptsize Regression with learned metric} \\[0.5em]
        \includegraphics[width=0.49\textwidth, trim=0 0 3cm 0, clip]{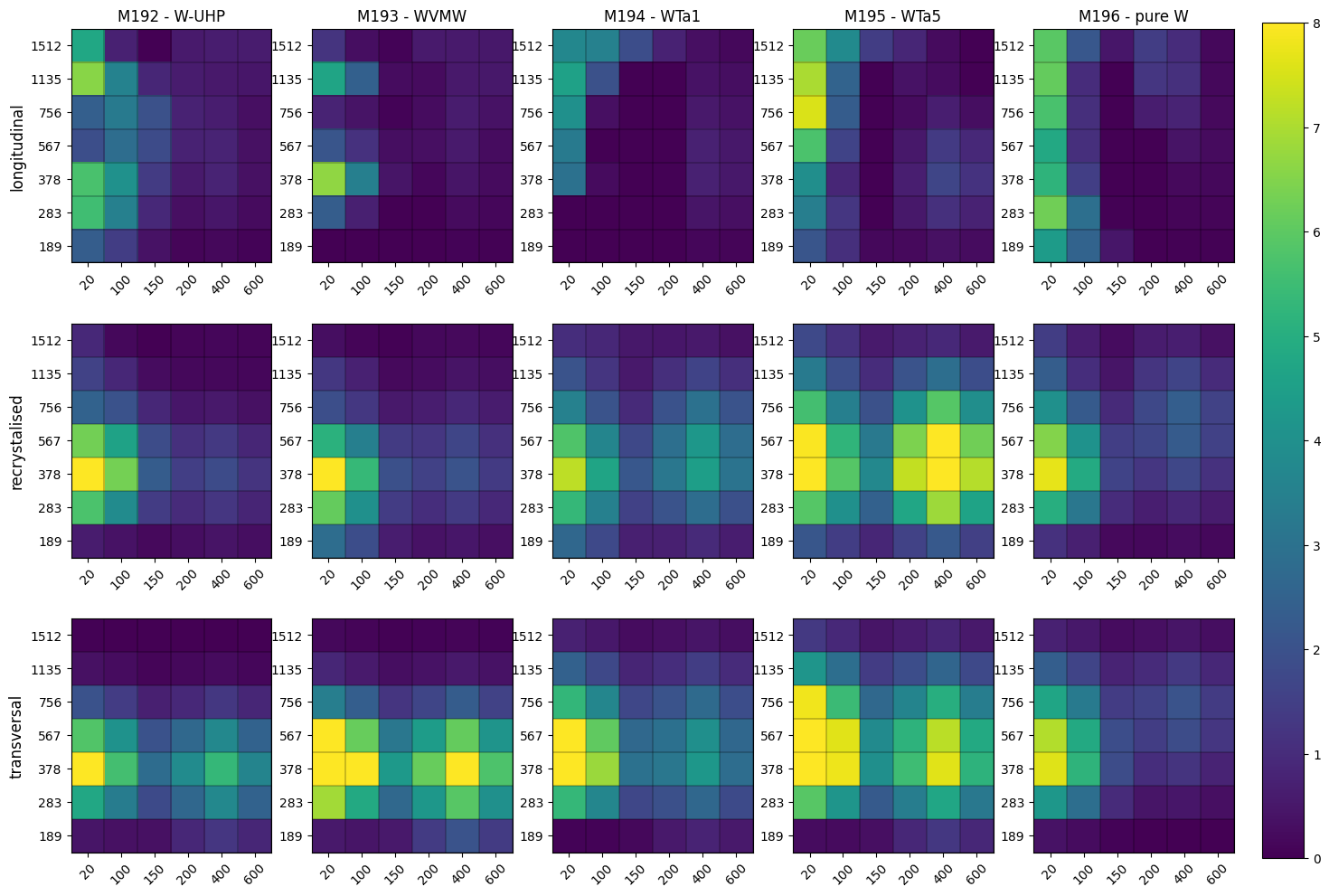}&
        \includegraphics[width=0.49\textwidth, trim=0 0 3cm 0, clip]{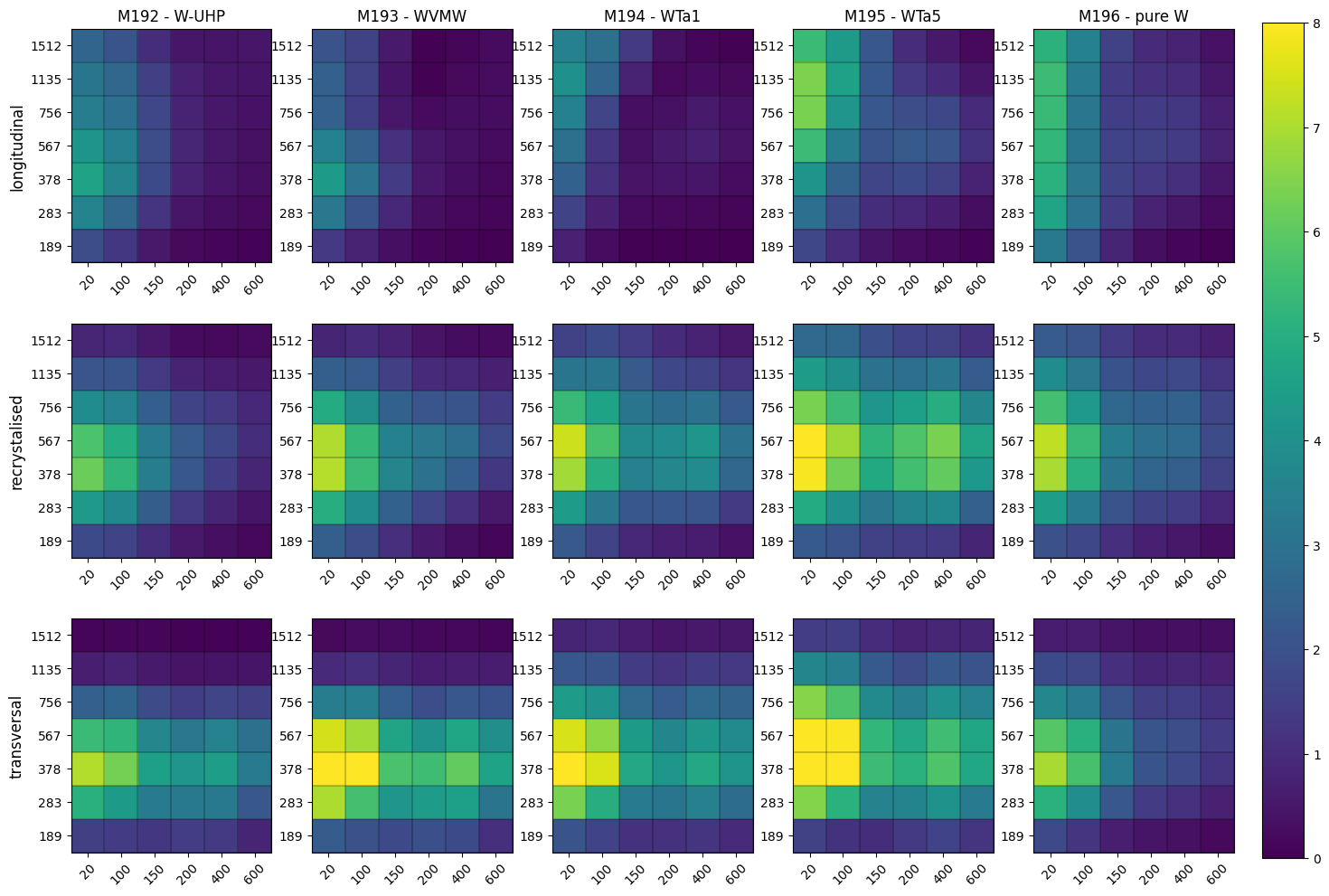}&
    \end{tabular}
  \caption{Top row: Experimental test condition matrix across microstructure, alloy, base temperature, and heat flux conditions. 
  Bottom row: Predicted crack density across the same parameter space using kernel ridge regression with (left) raw input parameters and (right) the learned metric.}
\label{fig:mat:crack_density}
\end{figure}


These differences are further evident in the spatial structure of the predictions (Fig.~\ref{fig:mat:crack_density}). 
While the input-based model closely reproduces the training data across the parameter grid, the learned-metric model yields smoother and more coherent predictions, interpolating across experimental conditions in a manner consistent with morphology-based similarity.

Together, these results highlight that meaningful generalization requires a notion of similarity that reflects observed surface morphology rather than raw proximity in parameter space. 
In the present approach, this is achieved by learning a metric (and corresponding kernel) that aligns parameter-space distances with distances in the diffusion feature space.
A key aspect of this formulation is that the metric is learned \emph{independently} of any specific target quantity. 
As a result, the learned kernel captures how variations in material properties and exposure conditions translate into differences in damage morphology, independent of a particular scalar label.

This separation between representation learning and prediction offers several advantages. 
First, once the metric is learned, it can be reused to predict \emph{any} quantity derived from the images, such as crack density, crack length, or connectivity measures, without retraining the underlying similarity structure. 
Second, it enables efficient incorporation of new labels: as additional measurements become available, prediction models can be updated using the fixed kernel without recomputing the embedding or relearning the metric. 
Finally, learning the kernel from morphology rather than directly from labels improves robustness in data-sparse settings, as it leverages the full image dataset (including unlabeled structure) to define similarity between samples.

In this sense, the learned metric provides a data-driven notion of proximity in parameter space that reflects observed physical behavior, while kernel-based regression provides a flexible and scalable framework for predicting quantities of interest on top of this representation.

\paragraph{Limitations and data requirements.}
While the proposed framework enables robust, morphology-informed prediction, its performance is ultimately constrained by the coverage and informativeness of the available data. 
As shown in Fig.~\ref{fig:mat:overview}, the experimental test matrix is sparse and unevenly sampled, with limited representation of certain regimes, particularly transversal and recrystallised material cuts. 
This sparsity directly impacts predictive behavior: as observed in Fig.~\ref{fig:mat:parities}, the learned-metric model exhibits a systematic underestimation of crack density in the low-damage regime, while Fig.~\ref{fig:mat:crack_density} shows reduced fidelity in regions of the parameter space that are weakly sampled. For example, the highest flux predictions do not always show an increasing crack density due to unexplored parameter space, as would be expected.

Beyond data sparsity, the current input representation is limited in its ability to fully explain variability in the observed morphology. 
The input space is primarily defined by experimental conditions (e.g., base temperature, heat flux) and coarse material identifiers (alloy, cut), which do not capture many of the underlying physical factors governing crack formation. 
In particular, relevant descriptors may include microstructural characteristics (e.g., grain size distribution), as well as material properties such as thermal conductivity, elastic modulus, yield strength, crystallographic texture, and anisotropy. 
The absence of such information limits the ability of the model to distinguish between conditions that produce similar parameter values but different morphological outcomes.

This limitation is reflected in the results: even when the learned metric aligns parameter-space distances with morphology in diffusion space, residual discrepancies remain, indicating that variations in crack behavior cannot be fully explained by the available input variables alone. 
In this sense, the learned metric partially compensates for missing structure in the input space, but cannot fully recover information that is not explicitly represented.

Taken together, these observations suggest that improving predictive performance requires not only increased data coverage, but also richer and more physically informative input descriptors. 
Incorporating additional material properties and experimentally relevant features would enable a more faithful mapping between parameter space and observed morphology. 
Importantly, the current framework naturally accommodates such extensions: additional descriptors can be integrated into the parameter vector $p$, and the learned metric can adapt to identify which variables are most relevant for predicting damage. For example, computational E-beam testing described in the next section has been developed as an additional data source for the AI model. Physics simulation input and the other planned improvements are detailed for the June demonstration described in Section \ref{sec:june}.

\subsection{Physics modeling to enhance MPEX AI Damage Assessment}
\label{sec:mat:physics}

Physics modeling has been a focus for MPEX AI Damage Assessment in order to support the AI predictions, providing mechanics and damage information where no experimental results exist. To accomplish this, improvements to the existing physics simulation tools for crack prediction were necessary, followed by validation with experimental results, and finally demonstration of the improved E-beam physics prediction, to be used within AI surrogate-driven automated workflows.

The ORNL-developed CabanaPD fracture mechanics software has been updated and applied to thermomechanically driven failure~\cite{cabanapd}. CabanaPD employs a peridynamics framework that directly simulates crack initiation and propagation and is open source: \url{https://github.com/ORNL/CabanaPD}.
Peridynamics (PD) is a nonlocal reformulation of classical continuum mechanics naturally suited for the simulation of material cracking and failure~\cite{SILLING2000, SILLING2007}. Unlike traditional continuum mechanics, which struggles to represent discontinuities in materials, PD uses integro-differential equations to accommodate cracks and other damage. Furthermore, CabanaPD uses a meshfree approach~\cite{SILLING2005}, which overcomes the challenges of mesh-based methods associated with constraining crack propagation to the underlying mesh, making CabanaPD highly effective in capturing advanced fracture phenomena. PD has been demonstrated in the literature to correctly produce the formation and propagation of complex cracks induced by thermal shocks~\cite{GIANNAKEAS20183037, DANTUONO201774}. Although it is a computationally expensive method, CabanaPD runs on both CPUs and GPUs across diverse hardware architectures, including the ORNL Frontier supercomputer, for scalable, high-resolution simulations through the Kokkos and Cabana libraries~\cite{kokkos2022,slattery2022}. CabanaPD has been shown to capture fracture due to thermomechanical shock loading in a previous project providing the basis for this work.

\subsubsection{Extraction of mechanical properties from experiments}

In order to make physics-based predictions for the same E-beam conditions as in Section~\ref{sec:mat:ai}, the published mechanical properties were extracted~\cite{wirtz2013, wirtz2016, linke2011} and input into the relevant CabanaPD model forms. Notably, the published stress-strain responses required a machine-compliance correction, using separate acoustically measured elastic moduli, to ensure consistent and accurate mechanical data. Data is available for all transverse and longitudinal samples at 300, 500, and 1000~$\degree $C; however, for recrystallized samples, only pure W curves were published. 

\begin{figure}
  \centering
  \includegraphics[scale=0.15]{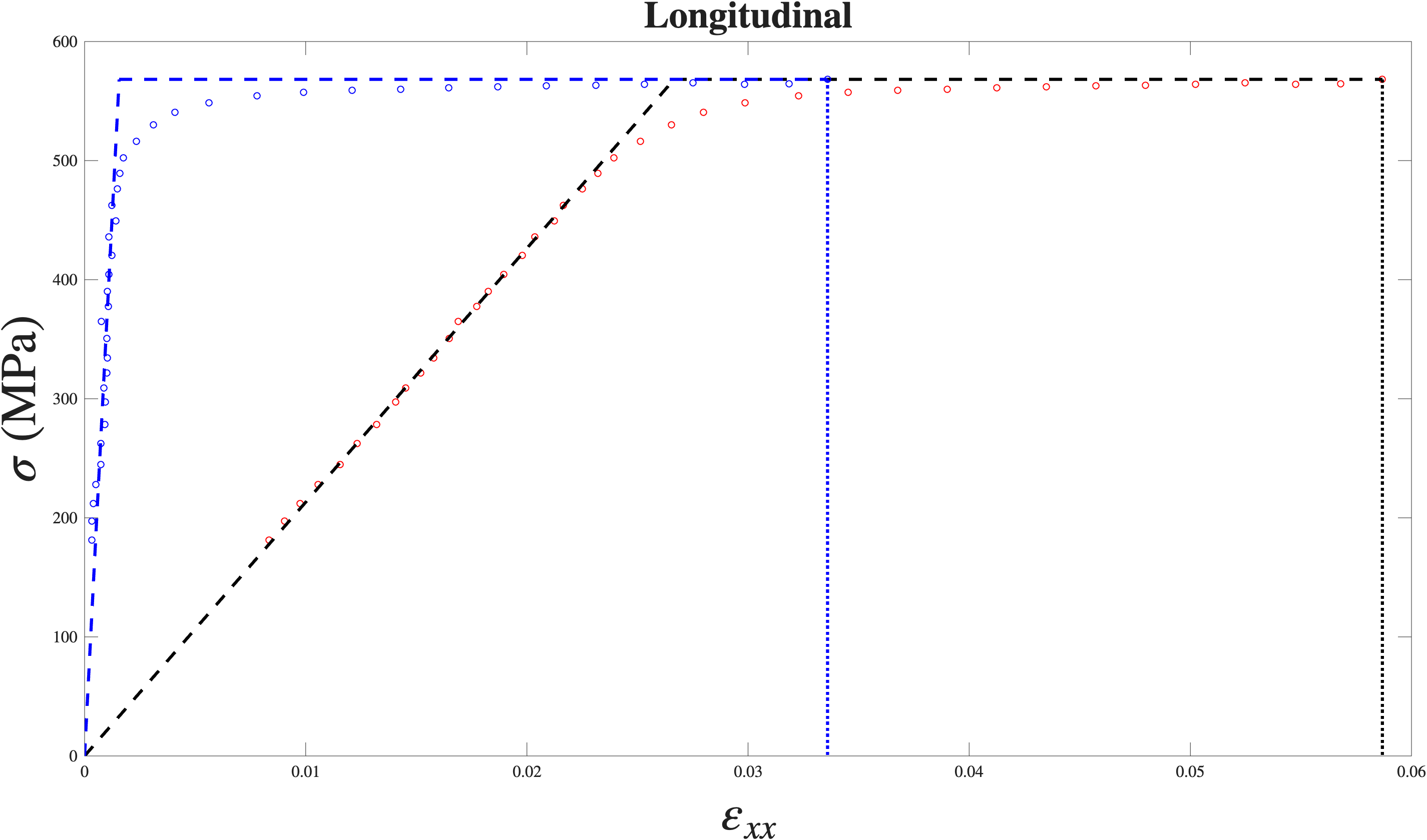}
  \hspace*{0.2in}
    \includegraphics[scale=0.15]{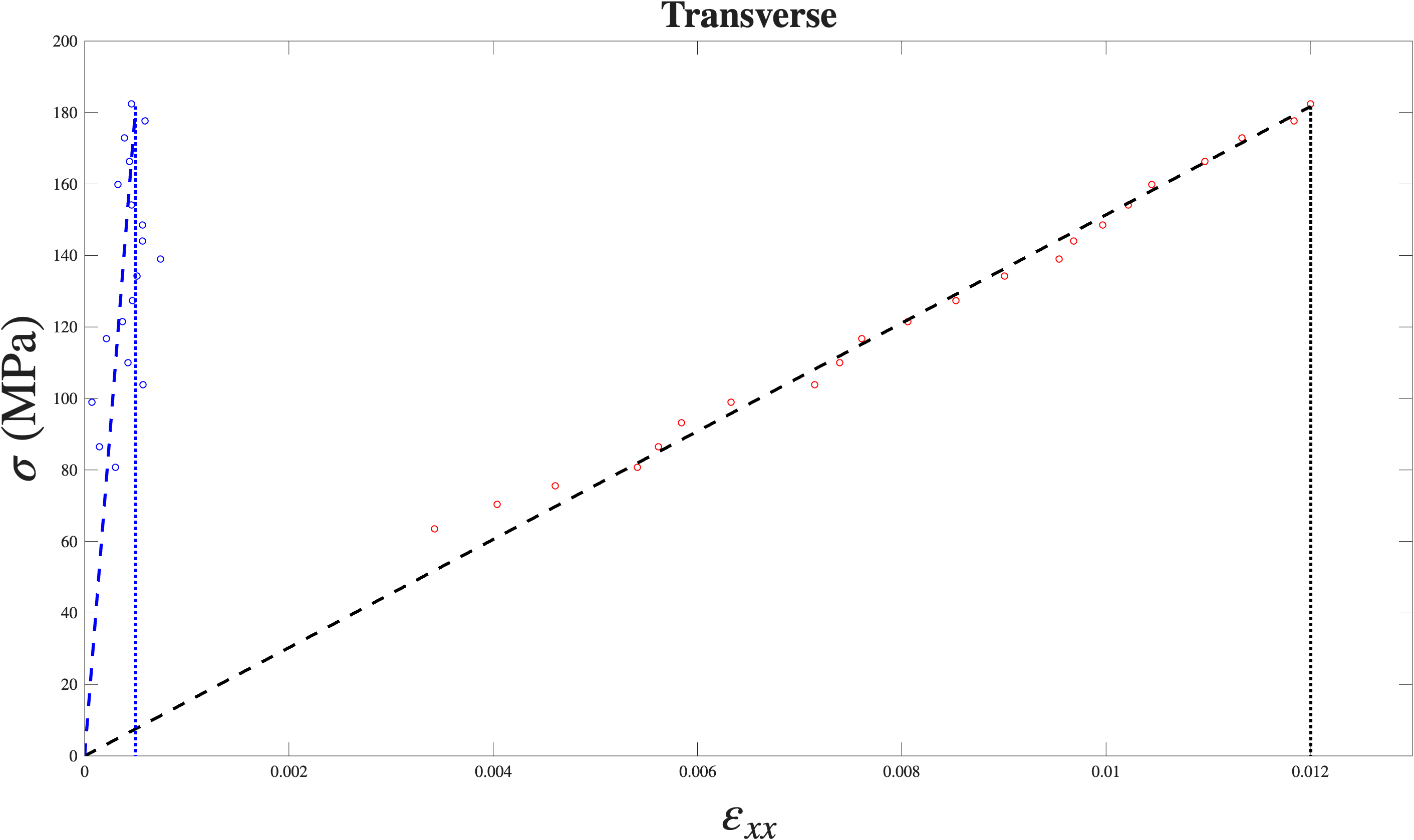}
  \caption{
  Experimental stress-strain data with model parameterizations for longitudinal and transverse orientations for WUHP at 300 $\degree$C~\cite{wirtz2013}. The red points show the original data and blue points the  machine-compliance-corrected data. The dashed lines show the elastic and elastic-perfectly plastic model fits.}
\label{fig:mat:stress-strain fit}
\end{figure}

The elastic modulus was fit as a function of temperature for the five different grades of tungsten, for both as-received and recrystallized samples, using data from~\cite{wirtz2013}. This enables direct input for CabanaPD and also calculation of the machine-compliance correction.
The experimental data for the five tungsten grades at each available temperature were digitized, and a fit was performed using an elastic or elastic–perfectly plastic model as appropriate. The separately measured moduli was used to correct only the elastic portion of the curve. This process is illustrated for a single tungsten grade (WUHP) at 300 $\degree$C, for both the longitudinal and transverse orientations in Fig.~\ref{fig:mat:stress-strain fit}. The elastic modulus, yield stress and strain, and failure strain model fits comprise the full set of material parameters used in the CabanaPD simulations.

\subsubsection{Expanded model fidelity}

\paragraph{Material anisotropy}
Prior to the MPEX AI Digital Twins project, CabanaPD had the basic thermomechanical simulation capabilities, including heat transfer, elastic and plastic deformation, and temperature boundary conditions. 
The code supports a range of models with varying levels of complexity, but in this work we rely on bond-based models to balance computational cost and simulation accuracy.  
Missing from CabanaPD was the effect of microstructure, which was critical to the previous experimental results and predicted by the MPEX AI Damage Assessment Digital Twin. While direct representation of the microstructure is a future goal of the project, first CabanaPD was extended to represent the continuum-level anisotropy arising from the underlying microstructure.

\begin{figure}
  \centering
  \includegraphics[scale=0.5]{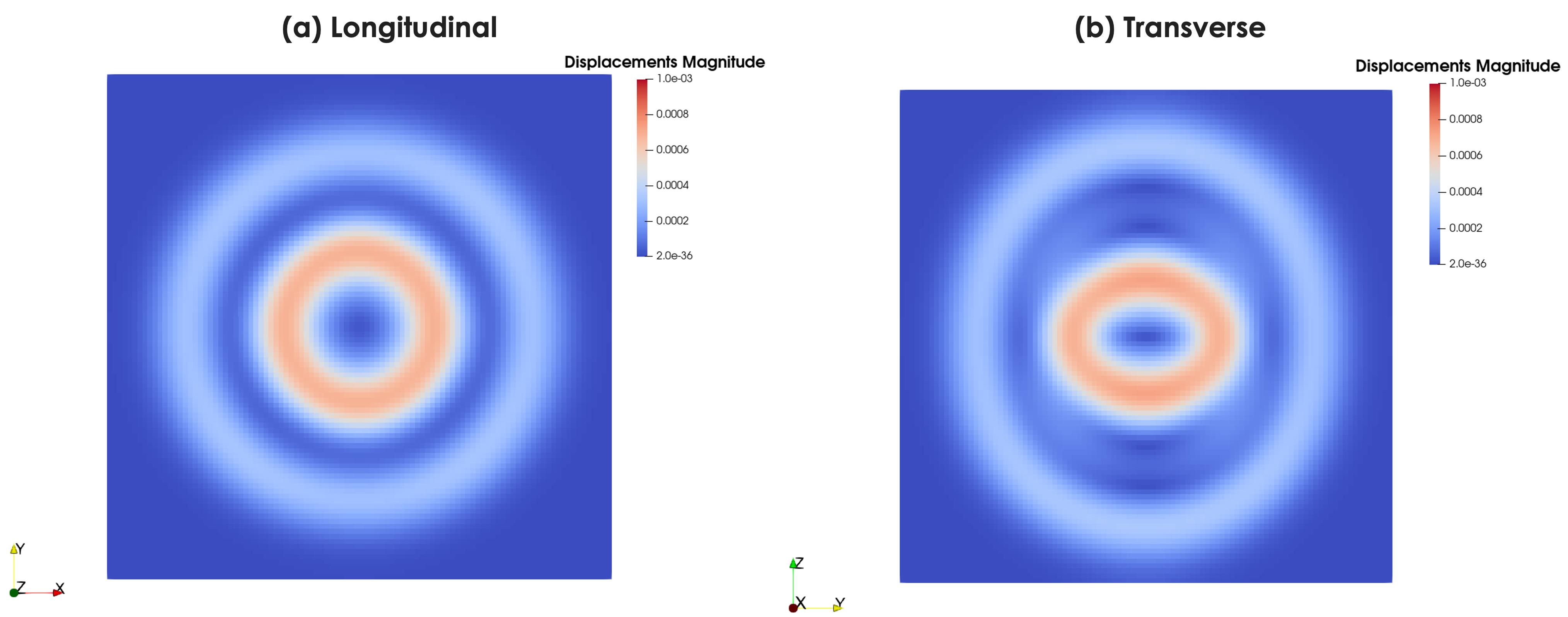}
  \caption{Anisotropic wave propagation for the transversely isotropic forged tungsten microstructures of interest.}
\label{fig:mat:aniso}
\end{figure}

For the tungsten alloys of interest here, the recrystallized alloys create a statistically isotropic system, while the rolled samples result in transversely isotropic mechanical behavior. CabanaPD was extended to include this, following Trageser and Seleson~\cite{trageser2019}, to produce differing mechanics in the transverse and longitudinal directions. 
The underlying orientation dependence can be further used within more complex microstructures (with overall cubic or ultimately triclinic symmetries), as well as explicit microstructure representations. Elastic wave propagation is shown in Fig.~\ref{fig:mat:aniso} to demonstrate this material anisotropy.

\paragraph{Long-time simulation}
An additional issue for direct numerical simulation of these materials is the disparity between the timescales of individual thermal pulses (on the order of $\mathrm{100~ms}$) and the many thousands of pulses of interest for material testing. Further exacerbating this issue is the computational time step size (dependent on system resolution; $<\mathrm{0.1~\mu s}$ for most cases). To achieve useful simulation, a quasi-static time-stepping approach was added to CabanaPD from the PD literature~\cite{wang2018}. The application of this approach to E-beam thermal shock simulation uses an alternating time-stepping scheme: standard explicit time integration during the largest thermal gradients, and quasi-static time integration to relax the system during the long temperature decay. This ensures accuracy where the most cracking is expected to occur and the ability to step over long time periods without damage.

\subsubsection{Physics model validation}
\begin{figure}
  \centering
  \includegraphics[scale=0.5]{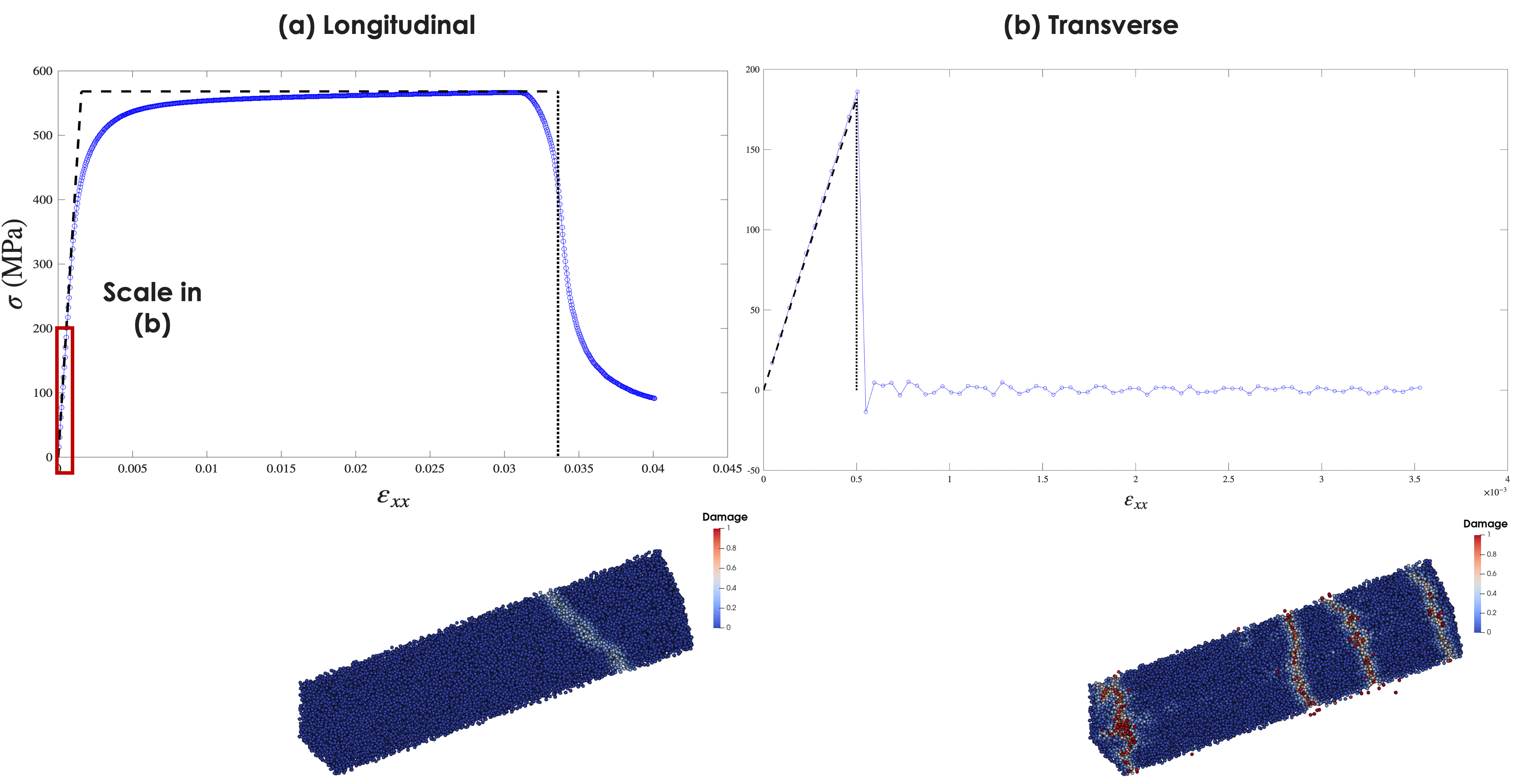}
  \caption{Stress-strain curves for ultra-high purity tungsten in the transverse and longitudinal directions. Dashed lines show the experimentally measured results simplified to an elastic–perfectly plastic model via a numerical fit. Insets show simulation result with particles colored by damage (relative number of broken bonds).}
\label{fig:mat:calibration}
\end{figure}

Using the converted mechanical response for the alloys of interest described above, CabanaPD model validation and calibration was undertaken through computational tensile tests following~\cite{wirtz2013}. CabanaPD models directly take as input mechanical properties and functional dependencies (e.g., temperature dependence), including the elastic constants, yield stress or strain, and failure strain. Due to discretization, surface effects, and strain-rate effects, the direct use of the experimentally measured mechanical properties for relatively small computational systems is not sufficiently accurate; therefore, a calibrated-parameter approach was developed. This ensures that, for a given discretization, the CabanaPD thermomechanical prediction reasonably match the experimental material behavior.  Fig.~\ref{fig:mat:calibration} shows the calibrated material response in both the longitudinal and transverse directions of WUHP at 300 $\degree$C as a representative example. While there is significant plasticity in the longitudinal direction, the transverse direction exhibits brittle elastic behavior. 
Convergence of all the mechanical properties as a function of computational system size and strain rate is under further investigation through model improvements to avoid further calibration needs. These results demonstrate the ability to accurately predict the range of mechanical responses necessary, at the continuum scale, to simulate accurate E-beam conditions shown in the following section.

\subsubsection{Crack prediction}

\begin{figure}
    \includegraphics[width=\textwidth]{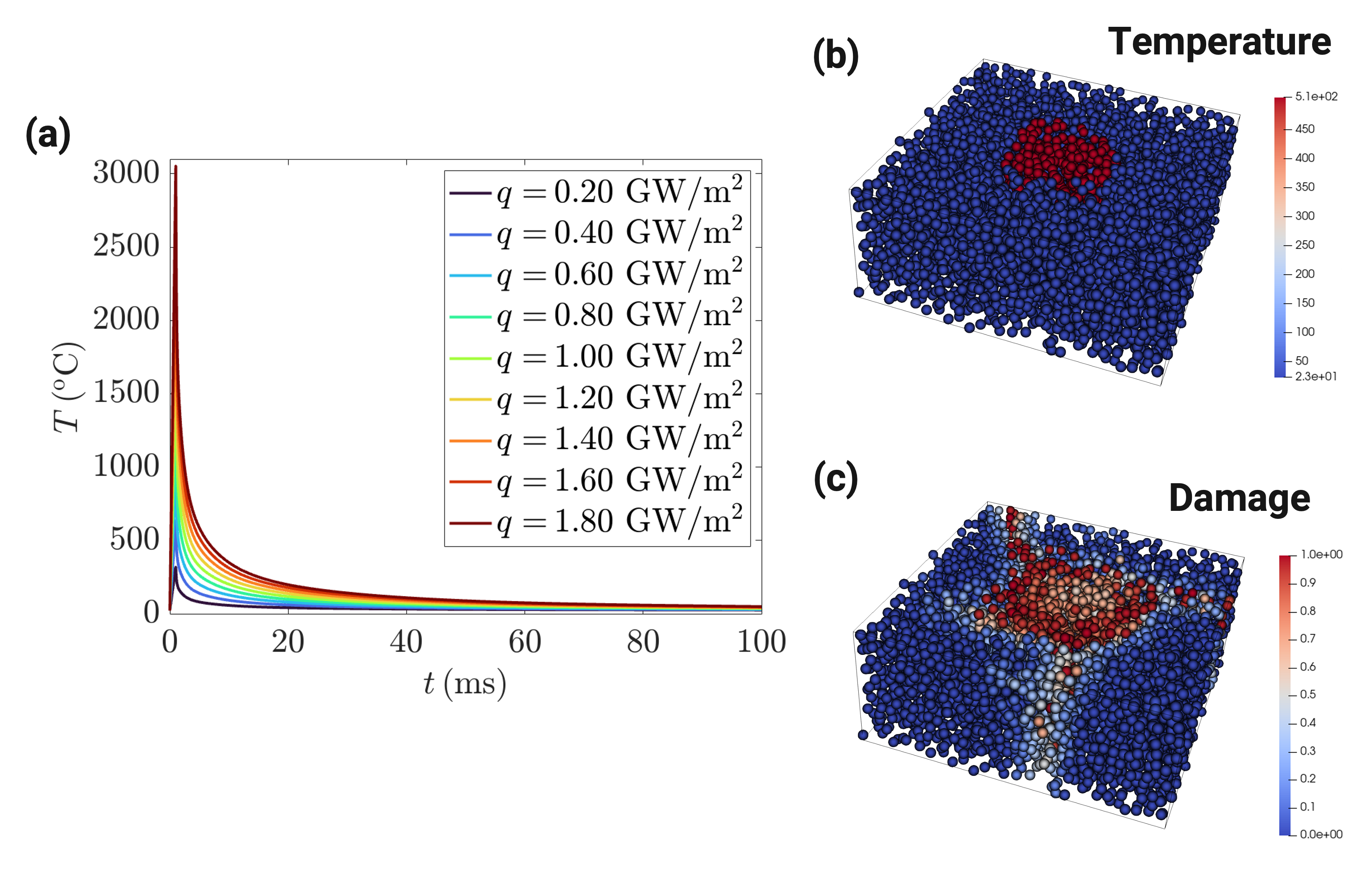}
    \caption{E-beam physics predictions with (a) imposed temperature profiles for CabanaPD matching experimental conditions, and a representative E-beam CabanaPD simulation for WUHP under a heat flux of $0.4~GW/m^{2}$ and a base temperature of $23\degree C$ showing (b) temperature upon damage and (c) cracking (damage indicating the relative number of broken bonds per particle).}
\label{fig:mat:shock}
\end{figure}

The computational E-beam predictions are the primary use of the CabanaPD physics simulation tool. Given the new developments described in the previous section, we are prepared to run high-throughput workflows of the same alloys, average microstructures, and E-beam conditions as dictated by the needs of the E-beam AI digital twin. Fig.~\ref{fig:mat:shock}a shows the varying thermal input conditions used in CabanaPD to match the experimental conditions of interest, as well as a single representative simulation for WUHP with ambient base temperature and moderate heat flux (0.4 GW/m$^2$). For the heat pulse affecting a local region of the material (see Fig.~\ref{fig:mat:shock}b) as in the experiment, this material fails (see Fig.~\ref{fig:mat:shock}c). We note that simulated damage is at the continuum scale without explicit microstructure detail, instead including the overall anisotropic mechanics of the combined microstructure. This result demonstrates the ability to predict the cracking and failure of the alloys of interest, including necessary continuum-scale anisotropic behavior, in order to support the AI model needs. Extraction of the top-surface damage will be used as input within the AI training process, as an analog of the SEM experimental results. Additional input forms, such as the full volumetric crack network, will also be considered. Predictions across the temperature and heat-flux space will be controlled by the planned sampling approach in order to dynamically update and retrain the E-beam AI Material assessment model for the upcoming June demonstration described in Section \ref{sec:june}.

\section{AUTOMATION AND WORKFLOWS FOR MPEX AI}
\label{sec:workflows}

Critical to the success and utility of the MPEX AI digital twins project is that the models developed are available and easy for MPEX users to leverage. This section details the work performed to develop customizable workflows for the AI and physics tools developed throughout the project.

Galaxy is a web-based data analysis platform that is widely used for bioinformatics and other data intensive science domains such as neutron scattering. It enables the integration of tools (roughly equivalent to numeric codes) with data and computational resources, managing provenance and ensuring reproducibility. Tools can be composed into workflows that can leverage computational resources spanning multiple sites. Workflows can be invoked via a web browser, custom user interfaces, or using scripts. Results can be visualized directly within the platform or using external visualization tools. We have established an instance of Galaxy at ORNL (\url{https://galaxy.ornl.gov}) in order to deploy MPEX tools and workflows. The ORNL instance has the capability of utilizing specific hardware resources (e.g. the fusion5 and fusion6 machines), internal cloud resources (via the ORNL Open Research Cloud), ORNL HPC resources (Frontier), and other DOE HPC resources (NERSC Perlmutter) from a single workflow.

\subsection{Customized workflows using Galaxy}

\subsubsection{Tool deployment in Galaxy}

To create a new software tool in Galaxy, two steps need to be completed. First, a build of the software that is capable of running on the target hardware best suited to execute it must be prepared. Second, an XML file is created that instructs Galaxy on how the tool operates, including defining the commands to run, the input files required for successful tool execution, and the output files that the tool creates. Additional optional metadata, such as dependency and help information, is also maintained in the XML file. Our Galaxy instance uses platform-independent containerized images for most tools. While this requires some additional work to build an image from the software source, it vastly simplifies dependency management and allows greater flexibility in choosing where the software will ultimately run, as it can run on any system that has a container engine installed. For HPC systems that don't allow the direct execution of containers, Galaxy has built-in support for batch job submission that can be utilized. In these cases, the software must be built directly on the target HPC system.

\begin{figure}
  \centering
  \includegraphics[scale=0.25]{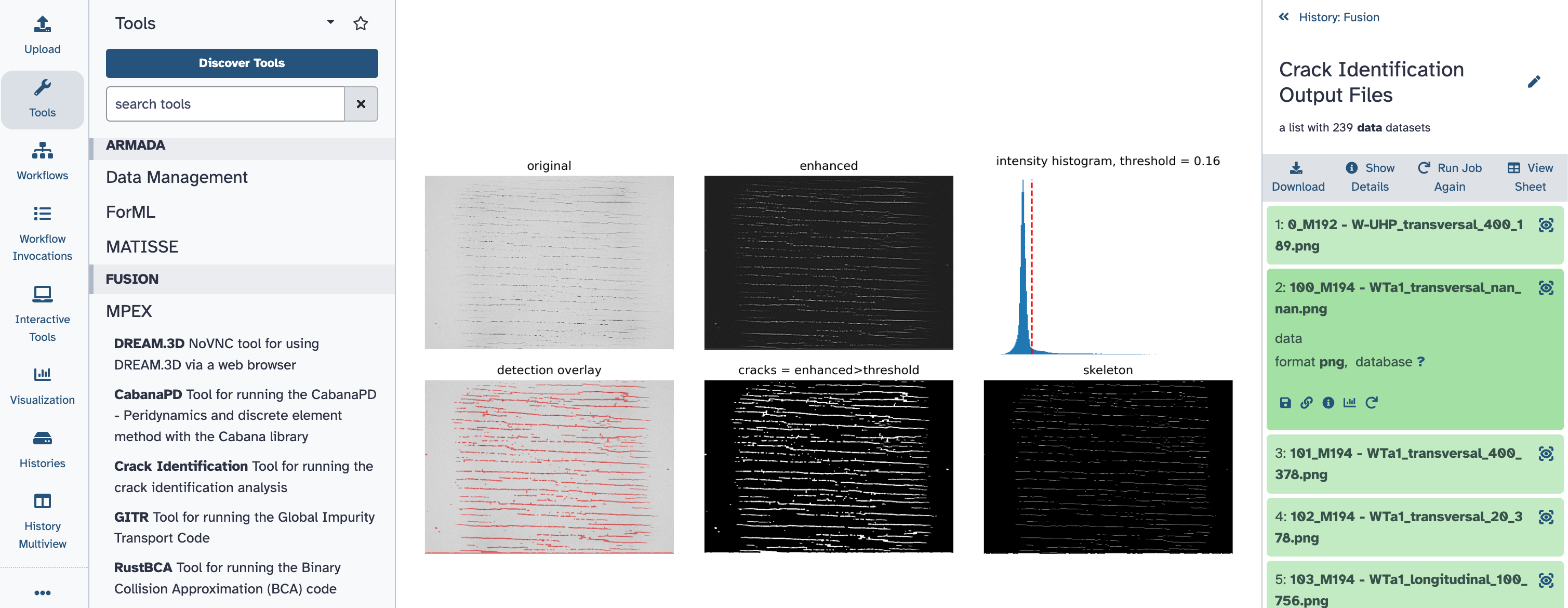}
  \caption{An example of a tool output that can be visualized directly in Galaxy. Here we visualize a PNG produced by an early prototype of the AI Crack Identification tool.}
\label{fig:galaxy:results}
\end{figure}

Once a tool is deployed to our Galaxy instance, it is accessible to the user to run. Users are able to import data files into Galaxy via an upload tool, using a script that interacts with the Galaxy API, or via (future) automated ingestion methods. Galaxy maintains a history of all imported data files, as well as tool executions and resulting output files created by the tool. These histories can be shared with others or used to create reproducible packages using techniques such as RO-Crates. Fig.~\ref{fig:galaxy:results} shows the Galaxy user interface with a selection of MPEX tools (left) and an example history (right). In addition, this figure shows that as well as managing tool execution and data file history, data files (inputs or outputs of a tool) can be visualized directly within the Galaxy user interface.

For the MPEX project, we have completed this process for a number of tools that will be used in plasma simulation, including DREAM.3D, GITR/GITRm, RustBCA, and CabanaPD. We have also completed a prototype of the MPEX AI Crack Identification tool described in \autoref{sec:mat:ai:cracks}.

\subsubsection{Workflow generation}

\begin{figure}
  \centering
  \includegraphics[scale=0.25]{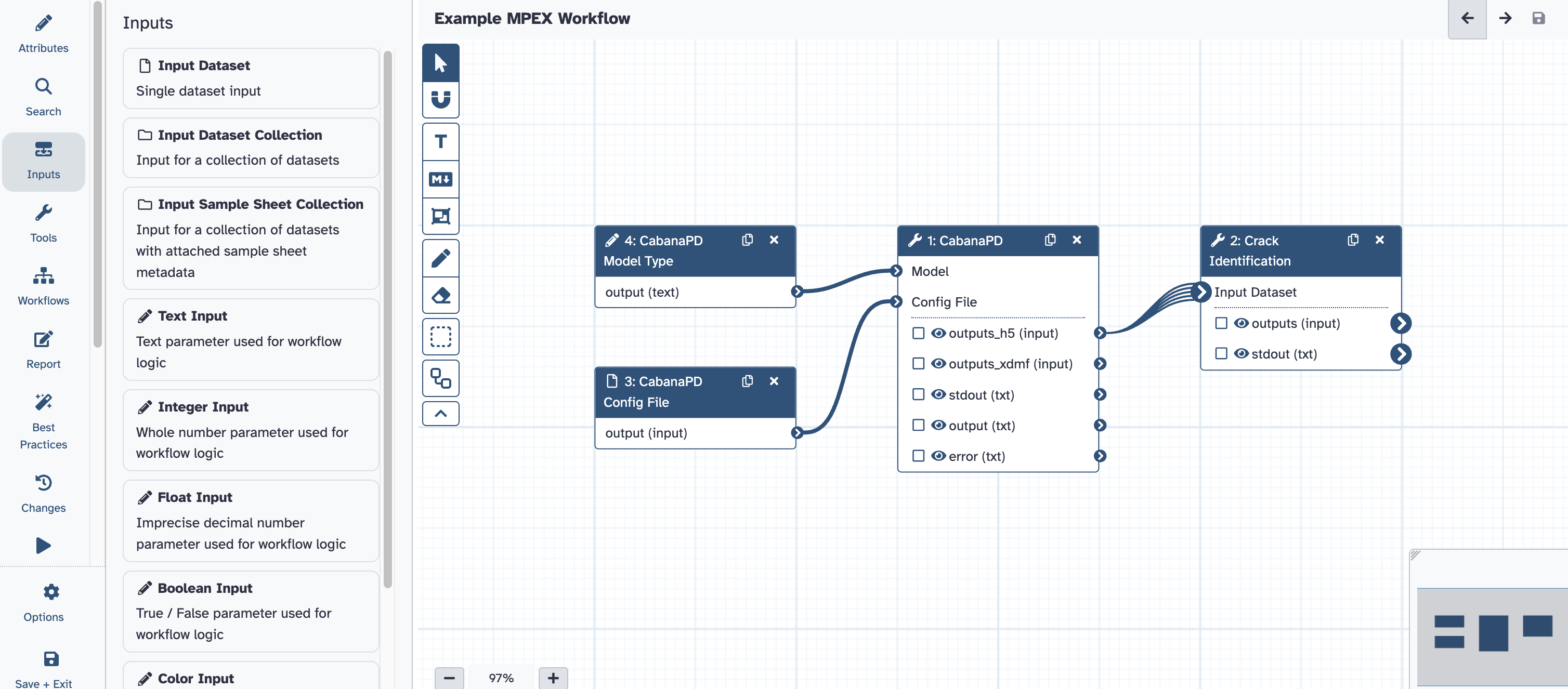}
  \caption{An example of the Galaxy workflow builder.}
\label{fig:galaxy:workflow}
\end{figure}

While users are able to run individual tools, the real power of Galaxy lies in it's ability to compose tools into workflows; a process that is analogous to composing library functions into a complex application. Galaxy workflows automatically manage the transfer of output data from one tool to the input of the next tool (or tools) in the workflows. When constructing a workflow, Galaxy will only allow tools with compatible inputs/outputs to be connected together, thus absolving the user from having to understand how to move data between the various software tools available to perform a wholistic analysis. Fig.~\ref{fig:galaxy:workflow} shows an example of the interface Galaxy provides for building these workflows. Workflows can be configured so that the user only needs to provide input files at the start of workflow execution, even though these input files may actually be inputs to multiple tools in the workflow. Once a workflow is created, it can be shared with other users on the system improving productivity and reuse.

Once the remaining tools in Fig.~\ref{fig:mat:workflow} have been created and integrated with Galaxy, our next goal will be to create a workflow that implements the full AI Material assessment digital twin. We then plan to build a custom user interface that will allow this workflow to be deployed to users without the need to interact with the Galaxy user interface (although this will remain an option for advanced users).

\section{JUNE DEMONSTRATION DELIVERABLES}
\label{sec:june}

The project is currently on track to complete all milestones for June demonstration. We now detail the current work towards these demonstrations in each project thrust.

\subsection{Helicon AI Hot Spot Digital Twin}
For the June milestone, the project will prioritize the following key developments:

\subsubsection{COMSOL--HERMES Two-Way Coupling}

Within the STRIPE workflow, HERMES-3 provides background plasma profiles for RF heating and PMI modeling, including density, temperature, and flow fields that determine RF absorption, collisionality, and neutral interactions. 

In the next phase, HERMES-3 and COMSOL will be iteratively coupled in a two-way framework. COMSOL will use plasma profiles from HERMES-3 to compute RF power deposition and sheath fields, while HERMES-3 will update plasma conditions based on the resulting RF heating. This iterative coupling is essential to obtain a plasma state that is self-consistent with RF wave propagation in Proto-MPEX and, ultimately, MPEX. This coupled workflow will be used to explore plasma density and helicon power levels to simulate the transition from edge to central RF heating. 

\subsubsection{C1: Reduced Fluid Plasma--Neutral Model Along Flux Tubes}

To enable efficient simulation and AI training, a reduced-order model based on the C1 code will be further developed and integrated. C1 solves coupled plasma and neutral fluid equations along individual flux tubes, making it well suited for use with the flux-tube mapping framework.

In this approach, the flux-tube mapper provides the parallel heat flux $Q_{\parallel}(l,m)$ as a source term along each field line. C1 then evolves plasma density, temperature, parallel flow velocity, and neutral species through ionization, recombination, and charge-exchange processes. This extends the framework beyond geometric heat-flux mapping by providing self-consistent plasma and neutral solutions along each flux tube.

This formulation captures key edge physics, including neutral recycling, collisional energy exchange, and flux-tube-dependent transport, while maintaining computational efficiency. As a result, C1 enables generation of large datasets across a wide parameter space for training surrogate AI models.

A further objective is to couple PICOS++ with C1 to incorporate kinetic corrections into the reduced-order framework. In this hybrid approach, the flux-tube mapper defines geometry and heat sources, C1 provides efficient fluid plasma--neutral transport, and PICOS++ captures non-Maxwellian effects and kinetic deviations along selected flux tubes.

\subsubsection{Coupling Strategy and Multi-Fidelity Framework}

The overall modeling strategy combines these components into a hierarchical multi-fidelity framework:
\begin{itemize}
    \item HERMES-3 provides high-fidelity 3D plasma background profiles.
    \item COMSOL computes RF power deposition and sheath physics.
    \item The flux-tube mapping framework translates heating into field-line-resolved transport quantities.
    \item C1 evolves plasma and neutral dynamics along flux tubes, enabling reduced-order modeling.
\end{itemize}

This multi-fidelity framework enables consistent modeling across scales, from detailed RF and plasma physics to reduced-order predictive models. It also provides a pathway for generating physics-based training data to support AI-driven hot-spot control and real-time optimization of MPEX operation.

\subsubsection{Integrated Multi-modal and Multi-Fidelity AI Framework}
We will expand our AI framework—currently integrating Proto-MPEX IR-camera measurements (Subsection~\ref{ssec:mmai})—to include the new Proto-Lite measurements and simulations from the multi-fidelity framework. This expanded effort will incorporate information from multiple sensors distributed across the Proto-MPEX-Lite facility. Our simulations and AI analyses have already influenced diagnostic upgrades, including the addition of new IR cameras viewing the downstream end of Proto-Lite MPEX (the “dump”), where our results suggest measurements can provide valuable information for understanding operation of the full device.

We will extend our variational autoencoder (VAE) into a “smart completion” model that infers the hidden operating state of MPEX from limited, incomplete diagnostics and provides actionable feedback for Proto-MPEX-Lite operations. We have completed an initial demonstration showing that a VAE trained on a single IR camera at the Helicon window, combined with flux-tube-resolved target simulations, can predict key heat-flux characteristics and help guide operations. Our current workflow is designed to scale to multi-fidelity simulation inputs, additional flux-tube-resolved target simulations, and multiple IR and experimental diagnostics.

We will train a conditional generative model using both paired and unpaired datasets so that it learns (1) how MPEX settings and physics parameters map to observable measurements, and (2) the statistical relationships between modalities (e.g., how patterns at the Helicon window relate to heat loads at the target). Low-fidelity simulations will provide broad coverage of operating space and teach coarse physics trends, while limited high-fidelity simulations and experimental measurements will anchor the model to more accurate behavior and to real diagnostic characteristics. The resulting model will generate self-consistent “device snapshots,” including predictions at unmeasured locations, synthetic signals for missing diagnostics, and associated uncertainty estimates.

Finally, we will continually update and validate the model throughout Proto-MPEX-Lite operations. For any trained model, we will produce predictions with uncertainty estimates and evaluate them against new experiments and simulations. As the dataset grows, we expect improved predictive accuracy and increased confidence in the uncertainty quantification.   

\subsection{E-beam AI Damage Assessment Digital Twin}

The primary milestone objective for the material assessment component of the project is to demonstrate learning the crack density map across alloys with minimal sampling of combined experimental and simulation data. The baseline AI crack prediction tool has been shown to accurately predict material response in regions where experimental data is available, but does not have sufficient information to predict across the entire input space (see \autoref{sec:mat:ai}). The CabanaPD thermomechanical fracture simulation tool has been successfully verified and validated in order to fill gaps in the available experimental-only data (see \autoref{sec:mat:physics}). The planned sampling approach is described in the following section and will be driven by the expected AI model improvement for each potential simulation of material and process condition to reduce uncertainty and refine the expected threshold for cracking. This will notably leverage the automated workflows developed for the project described in Section \ref{sec:workflows}.

\subsubsection{Gaussian processes for targeted material search.}
The kernel-based regression framework described in Section \ref{sec:mat:damage} can also be interpreted in a probabilistic setting as a Gaussian process (GP) model.
Specifically, by treating the learned kernel $K(p_i, p_j)$
as a covariance function, we obtain a Gaussian process prior over functions defined on the parameter space. 
In this view, kernel ridge regression corresponds to the posterior mean of a Gaussian process with Gaussian observation noise, providing a probabilistic interpretation of the predictor.

This perspective offers several advantages for the upcoming demonstration of the E-beam AI digital twin. 
First, it provides a natural mechanism for \emph{uncertainty quantification}, allowing prediction variance to be estimated alongside the mean. 
This is particularly important in the present setting, where the dataset is sparse and unevenly sampled (Fig.~\ref{fig:mat:overview}), and predictive confidence should reflect the local density of available data. 
Second, the GP formulation highlights the role of the learned metric: by defining the kernel through $C^{-1}$, the covariance structure is directly informed by similarity in the diffusion feature space, ensuring that uncertainty and interpolation are governed by morphology-consistent relationships. 
Finally, the probabilistic interpretation enables principled extensions, such as incorporating prior knowledge, handling noisy or partial observations, and performing Bayesian updates as new data become available. This formulation will enable input of physics simulation, as well as additional external data in future work.
In this way, the learned metric and kernel not only support deterministic prediction via kernel regression, but also define a flexible probabilistic surrogate model for damage prediction across the experimental parameter space.

The Gaussian process formulation naturally enables a strategy for exploring the material and process parameter space. 
Given the GP posterior mean and variance defined by the learned kernel, one can define acquisition functions that balance exploitation of regions with high predicted damage and exploration of regions with high uncertainty. 

In the present context, the key advantage is that the acquisition strategy is informed by the learned metric, which encodes similarity in terms of observed surface morphology. 
As a result, exploration in parameter space is guided not only by proximity in experimental conditions, but by proximity in the induced feature space that reflects damage mechanisms. 

Formally, given an acquisition function $a(p)$ derived from the GP posterior, the next experiment can be selected as
\begin{align*}
    p^\ast = \arg\max_{p \in \mathcal{P}} a(p),
\end{align*}
where $\mathcal{P}$ denotes the admissible parameter space. 
This provides a principled framework for sequential experimental design, enabling efficient identification of regions associated with high damage, transitions in behavior, or poorly understood regimes. 
In this way, the learned metric and GP surrogate together support not only prediction, but also adaptive exploration of the material design space.

\subsubsection{Integrated Multi-Fidelity E-beam AI Damage Assessment}

The planned sampling approach will use a combination of available experimental data and the expanded simulation capabilities from \autoref{sec:mat:physics}. This updated, multi-fidelity AI model will require an embedding of the simulation cracking result, just as with the combined experimental SEM images and process conditions, for a cohesive representation of the state space. This will primarily include images of the damaged surface from CabanaPD simulations as the computational analog of the SEM results. Our workflow will thus dispatch simulations as needed to fill the process and material space and make improved E-beam AI predictions of crack density.

One additional near-term extension of the E-beam AI surrogate is to directly include material mechanical property metadata for model training in order to generalize predictions to other material classes. Current predictions include alloy and microstructure identifiers, but need additional material properties for enable predictive interpolation beyond the current data.
The aforementioned extensions to the AI model inputs through sampling of experimental and physics simulation crack predictions will complete the June milestone demonstration for the MPEX AI Damage Assessment Digital Twin.

\printbibliography

\end{document}